\documentclass[]{aa}  
\usepackage{graphicx, color}

\newcommand{\cd}{d$^{-1}$\,}
\newcommand{\kms}{km~s$^{-1}$\,}

\newcommand{\vsini}{$v\sin{i}$\,}

\newcommand{\ignore}[1]{}
\usepackage{txfonts}
\begin{document}
   \title{Discovery of non-radial pulsations in the spectroscopic binary Herbig Ae star RS Cha
   \thanks{Based on observations collected at the 1m McLellan telescope at Mt John, NZ
   }
}


   \author{T. B\" ohm
          \inst{1} 
           \and
          W. Zima\inst{2}
          \and
          C. Catala\inst{3}
          \and
          E. Alecian\inst{4}
           \and
          K. Pollard\inst{5}
           \and
          D. Wright\inst{5}
          }

   \offprints{T. B\"ohm}

   \institute{Laboratoire Astrophysique de Toulouse-Tarbes, Universit\'e de Toulouse, CNRS, 14 avenue Edouard Belin, F - 31400 Toulouse, France. \email{boehm@obs-mip.fr}
       \and
          Instituut voor Sterrenkunde, KU Leuven, Celestijnenlaan 200D, B-3001 Leuven, Belgium
       \and
          LESIA, Observatoire de Paris-Meudon, 5 place Jules Janssen, F - 92195 Meudon
       \and
           Dept. of Physics, Royal Military College of Canada, PO Box 17000, Stn Forces, Kingston, Canada K7KK 7B4 
       \and
           Dept. of Physics and Astronomy, University of Canterbury, Private Bag 4800, Christchurch, New Zealand 
                             }

   \date{Received December 2, 2008; accepted December 15, 2008}

 
  \abstract
   {To understand the origin of stellar activity in pre-main sequence Herbig Ae/Be stars and to get a deeper insight in the interior of these enigmatic stars, the pulsational instability strip of Palla and Marconi is investigated. In this article we present a first discovery of non radial pulsations in the Herbig Ae spectroscopic binary star RS Cha.}
   {The goal of the present work is to detect for the first time directly by spectrographic means non-radial pulsations in a Herbig Ae star and to identify the largest amplitude pulsation modes.}
   {The spectroscopic binary Herbig Ae star RS Cha was monitored in quasi-continuous observations during 14 observing nights (Jan 2006) at the 1m Mt John (New Zealand) telescope with the Hercules high-resolution echelle spectrograph. The cumulated exposure time on the star was 44 hrs, corresponding to 255 individual high-resolution echelle spectra with $R = 45000$. Least square deconvolved spectra (LSD) were obtained for each spectrum representing the effective photospheric absorption profile modified by pulsations. 
Difference spectra were calculated by subtracting  rotationally broadened artificial profiles; these residual spectra were analysed and non-radial pulsations were detected. A subsequent analysis with two complementary methods, namely Fourier Parameter Fit (FPF) and Fourier 2D (F2D) has been performed and first constraints on the pulsation modes have been derived. }
   {For the very first time we discovered by direct observational means using high resolution echelle spectroscopy non radial oscillations in a Herbig Ae star. In fact, both components of the spectroscopic binary are Herbig Ae stars and both show NRPs. The FPF method identified 2 modes for the primary component with (degree $\ell$, azimuthal order $m$) couples ordered by decreasing probability:   $f_1$ = 21.11 \cd with ($\ell$,$m$) = (11,11), (11,9) or (10,6) and  $f_2$ = 30.38 \cd with ($\ell$,$m$) = (10,6) or (9,5). The F2D analysis indicates for $f_1$ a degree $\ell$ =  8-10. For the secondary component, the FPF method identified 3 modes with ($\ell$,$m$) ordered by decreasing probability:   $f_1$ = 12.81 \cd with ($\ell$,$m$) = (2,1) or (2,2),  $f_{\mathrm 2b}$ = 19.11 \cd with ($\ell$,$m$) = (13,5) or (10,5) and  $f_3$ = 24.56 \cd with ($\ell$,$m$) = (6,3) or (6,5). The F2D analysis indicates for $f_1$ a degree $\ell$ =   2 or 3, but  proposes a contradictory identification of $f_2$ as a radial pulsation ($\ell$ = 0).
}
   {}


   \maketitle
%

\section{Introduction}

Asteroseismology represents a modern tool for studying the stellar interiors of the enigmatic group of pre-main sequence (PMS) stars of intermediate mass ($\sim$ 2-8 M$_{\odot}$),  so called Herbig Ae/Be stars (\cite{Herbig};  \cite{Strom}; \cite{fimu}; \cite{fija}). Since their first systematic classification in 1960 (\cite{Herbig}) this group of young stars has been extensively studied, but one of the major questions remains unanswered: how can the intense stellar activity and strong stellar winds (\cite{praderie};  \cite{catala}; \cite{caku}; \cite{tba6} and  \cite{tba8}), as well as the many highly variable emission lines observed in the spectra of these stars be explained?  

Magnetism is frequently invoked as being responsible for active stellar phenomena. 
The position of the Herbig stars in the HR diagram indicates that they are in the 
radiative phase of their contraction towards the main sequence (\cite{iben};
\cite{gilliland}), and should in principle not possess any outer convective 
zone. In its absence, the classical magnetic dynamo mechanism could not be responsible for the
observed active phenomena. Moreover, recent spectropolarimetric observations of Herbig stars indicate the presence of significant large-scale magnetic fields only in a small fraction of them (\cite{tba16}, \cite{tba18} and \cite{alecian2008}). This fraction, close to 10\%, is consistent with a magnetic field distribution similar to that of the main sequence stars. In that scenario, magnetic Herbig Ae stars would be the progenitors of main sequence Ap/Bp stars. These recent results are in agreement with the primordial fossil field hypothesis.
Unless magnetic fields  in Herbig Ae/Be stars are of very complex nature and can therefore not be detected by spectropolarimetric means, in opposition to clearly organized dipolar fields, the pure magnetic origin of activity in these stars is more than doubtful. 

For many years non-classically generated magnetic fields have been used as a potential explanation for the observed active phenomena. 
Even if the presence of complex fields can not be ruled out it is of major importance to investigate other possible external or internal origins of this tremendous amount of dissipated energy, as witnessed by chromospheres and coronae, but also variable spectral lines, winds and bipolar jets. 

The only way of studying in detail the internal stellar structure is the analysis and the modelling of stellar pulsations, if observed. As an example,
PMS stars gain their energy from gravitational contraction and therefore differ signiÞcantly from post-main-sequence stars having already 
processed nuclear material - stellar pulsations are sensitive to these differences of internal stellar structure (see e.g. \cite{suran}). As of today,  the internal stellar structure of PMS stars is not yet well constrained. 
Since few years the existence of pulsating intermediate mass PMS stars is known (\cite{breger1972}; \cite{kuma}; \cite{donati}). This observational result motivated Marconi \& Palla (\cite{mapa}) to investigate the  pulsation characteristics of HR\,5999 theoretically, 
which enabled them to predict the existence of a pre-main-sequence 
instability strip, which is being crossed by most of the intermediate mass PMS objects for a significant fraction of their evolution to the main sequence. This strip covers approximately the same 
area in the HR diagram as the $\delta$ Scuti variables. Zwintz (\cite{zwintz08}) compared, based on photometry, the observational instability regions for pulsating pre-main sequence and classical $\delta$ Scuti stars and concluded that the hot and cool boundaries of both HR diagram instability regions seem to coincide. This preliminary result deserves further study by aim of full asteroseismological approach based on spectroscopy.

As of today, more than 30 intermediate-mass PMS stars have revealed to be pulsating at time-scales typical of $\delta$ Scuti stars (see e.g. \cite{kuma}; \cite{kuca}; \cite{donati};  \cite{tba15};  \cite{marconi2002};  \cite{rima}; \cite{zwintz}; \cite{ca} and references therein). 

RS Chamaeleontis is a bright spectroscopic eclipsing binary star. Both components are Herbig Ae PMS stars of similar mass (close to 1.9 M$_{\odot}$).  Recently the age of RS\,Cha has been determined to 6$^{+2}_{-1}$\,Myr  (\cite{luhman}), which verifies it's PMS nature.
\cite{andersen} already reported small amplitude radial velocity variations on top of the binary radial velocity curve for both components of RS Cha, suggesting the possible presence of stellar pulsations. Photometric observations by \cite{mcinally} revealed short-term variations in at least one of the two components, possibly linked to stellar pulsations. Very recently, Alecian et al. (\cite{alecian2005}) reported radial velocity variations  in the residual velocity frame (cleaned for orbital velocity) with amplitudes up to a few \kms\ and periods of the order of 1h, indicative of $\delta$ Scuti type pulsations. 

The aim of our study of the two components of RS Cha is to provide a first set of asteroseismic constraints for forthcoming
non-radial pulsation models by determining unambiguously a higher number of periodicities and
identifying, in a second step, the corresponding pulsation modes with their respective degree $\ell$ and azimuthal number $m$. 

To achieve this goal,  we decided to perform high resolution spectroscopic observations on a large time basis and with optimized time coverage. 

Section \ref{previous} reviews previous related work, Sect.
\ref{observations}  describes the observations and data reduction, Sect. \ref{orbit}  summarizes results of the
orbit determination, Sect. \ref{detection} reveals the detection of non-radial pulsations in both components of RS Cha,
Sect. \ref{pulsaprim}  and \ref{pulsasecond}  present frequency analysis and moment identification in the primary and secondary component, respectively. A discussion is proposed and a conclusion is drawn in Section \ref{dissconc}.


\section{Previous related work}
\label{previous}

The pre-main sequence spectroscopic eclipsing binary RS Cha has been studied extensively throughout the last years. Thanks to its eclipsing nature and the known inclination angle the system has fully been calibrated (\cite{alecian2005}, \cite{alecian2007}, \cite{alecian2007b}).
Table \ref{tab:para} summarizes the main results.

\begin{table}
\caption[]{Parameters of RS\,Cha. References: [1]  \cite{alecian2005}, 
[2] \cite{ribas}, [3] \cite{clausen}.}
\begin{tabular}{lccc}
\hline\hline
Parameter    &     Primary  & Secondary & References\\
\hline
$M/\mathrm{M}_\odot$ & 1.89$\pm$0.01 & 1.87$\pm$0.01 & [1]   \\
$R/\mathrm{R}_\odot$ & 2.15$\pm$0.06 & 1.87$\pm$0.01 & [1]   \\
$T_\mathrm{eff}~[\mathrm{K}]$ & 7638$\pm$76 & 7228$\pm$72 & [2]   \\
$\log(L/\mathrm{L}_\odot)$ & 1.15$\pm$0.09 & 1.13$\pm$0.09 & $L = 4\pi\,R^{2}\sigma\,T_\mathrm{eff}^{4}$    \\
$\log(g)~[\mathrm{cm\,s^{-2}}]$ & 4.05$\pm$0.06& 3.96$\pm$0.06 & $g = M\mathrm{G}/R^{2}$    \\
$v\,\sin\,i~[\mathrm{km\,s^{-1}}]$ & 64$\pm$6& 70$\pm$6 & [1]    \\
$P_\mathrm{orb}~[\mathrm{d}]$ & \multicolumn{2}{c}{1.67} 	 & [1]\\
$i~[\mathrm{\deg}]$ 	&  \multicolumn{2}{c}{83.4$\pm$0.3} & [3]\\
$\mathrm{[Fe/H]}$ & \multicolumn{2}{c}{0.17$\pm$0.01} & [1]\\
\hline
\end{tabular}
\label{tab:para}
\end{table}

\section{Observations}
\label{observations}

The analysis presented in this paper is based on a 14 nights observing
run in January 2006 at the 1m Mt John telescope equipped with the Hercules echelle spectrograph.  
We obtained quasi-continuous single-site observations of the target star during these 2 weeks and obtained a total of 
255 individual stellar echelle spectra, each spectrum having an individual exposure time of 10\,min. 
The star was observed in high resolution spectroscopy  at R $\approx$ 45000 and covering the wavelength area from 
457 to 704\,nm, spread over 44 orders. The detector was a 1kx1k  Site CCD.  The highest S/N (pixel$^{-1}$) values we obtained reached 210 on Jan 16th, corresponding to almost 300 per resolved element (2 pixels); typical values of S/N (pixel$^{-1})$ ranged around 80-150 in this run.
Table \ref{log} summarizes the log of  the observations.

\begin{table}
\caption[]{Log of the observations at Mt John Observatory, NZ, in Jan 2006. (1) and (2) Julian date (mean observation) (2,450,000+); (3) Number of high resolution RS Cha spectra; (4) typical range of S/N (pixel$^{-1}$) at 550\,nm (centre of V band) }
\begin{tabular}{ccccccc}
\hline\hline
 Date	& JD$_{\rm first}$	& JD$_{\rm last}$	& N$_{\rm spec}$	&S/N$_{\rm Range}$\\
(1)		&  (2)  			&  (3)        			&   (4)         		&   (5)       \\
\hline
Jan 09	& 3745.0101 		&				& 1	&   120 \\
Jan 10	& 3745.9544		& 3745.9733		& 2	&   80-100\\
Jan 12	& 3747.9537		& 3748.1779		&  28 & 70-90\\
Jan 13	& 3749.0013		& 3749.1930		&  23 & 100-120\\
Jan 14	& 3749.8911		& 3749.9911		&  12 &  70-90\\
Jan 15	& 3750.8921		& 3751.1919		&  31 &  150\\
Jan 16	& 3751.8958		& 3752.1970		&  29 &  90-120\\
Jan 19	& 3754.9015		& 3755.2006		&  33 &  100-130\\
Jan 20	& 3755.9245		& 3756.1993		&  25 &  90-120\\
Jan 21	& 3756.9068		& 3757.0386		&  14 &  60-150 \\
Jan 22	& 3757.9004		& 3758.1621		&  32 &  120-170\\
\hline
\end{tabular}
\label{log}
\end{table}
 
 The general observing strategy was to obtain as many 10 minute observations of the target star
 during the night as possible.  Before the beginning and after the end of the night we
 obtained  complete sets of calibrations: in general 6 bias images, 6 tungsten light flat field spectra, and, the closest possible to the 
 first stellar spectrum, 2 ThAr arc spectra (the calibration set at the end of the night being scheduled in the opposite order).
 In order to guarantee the possibility to monitor directly small spectral shifts we decided to 
 interleave regularly ThAr arc spectra during the night (every two hrs).
  
Most of the data reduction was carried out following standard reduction procedures using the ``ESPRIT'' 
spectroscopic data reduction package (\cite{donati}). This package also makes use of the ``optimal extraction algorithm" (Horne, \cite{horne}). Heliocentric velocity and Julian date correction was performed.
The intrinsic wavelength calibration accuracy achieved with the ``ESPRIT'' 2D-polynomial fit procedure is better 
than 0.22\,pm mean rms (i.e. 2.2\,m$\AA$, corresponding to 120\,ms$^{-1}$ at 5500\,$\AA$) for the Hercules data set.

The best way to correct the 2D-wavelength polynomial for evolution during the night due to the intrinsic
spectral instability of the instrument revealed to be the computation of least square deconvolved LSD profiles (\cite{donati}) of the more than 100 telluric water vapour lines contained in each stellar spectrum and to cross correlate them with respect to the first spectrum of the night. The precision of the above correction is around 50\,ms$^{-1}$ (see Appendix A in  \cite{tba9}). All spectra were then corrected with their individual time-dependent correction to the initial wavelength calibration based on the ThAr arc lamp spectrum acquired at the beginning of each night. More details about this procedure can be read in \cite{tba15}. 

The next step of the data reduction was to calculate for all 255 stellar spectra photospheric LSD-profiles,
using a mask corresponding to the spectral type A7, but taking care to eliminate from the line list all transitions showing non-photospheric behaviour (Hydrogen Balmer lines, He I D3, many FeII lines, Na I D,...). The equivalent photospheric profiles contain eventually multiplex information of 1930 individual photospheric lines, most of them being very weak lines. Typically, a spectrum of RS Cha with  S/N = 150 (pixel$^{-1}$) at 5500\,$\AA$ yields a typical S/N value of  1400 per velocity step (or pixel) in the resulting LSD profile. The accuracy of the radial velocity determination on an individual LSD profile scales, to the first order, with the inverse of the square root of the weighted multiplex gain and is therefore estimated to be as low as 12\,ms$^{-1}$; however, the limiting factor in the precision of the overall radial velocity correction is clearly the spectra-to-spectra cross correlation of telluric line LSD profiles. We therefore estimate the
finally achieved precision in radial velocity to be around 50\,ms$^{-1}$ for each individual stellar LSD profile of this data set, which is largely enough for our spectroscopic study.  

\begin{figure}[!ht]
  \includegraphics[width=9cm]{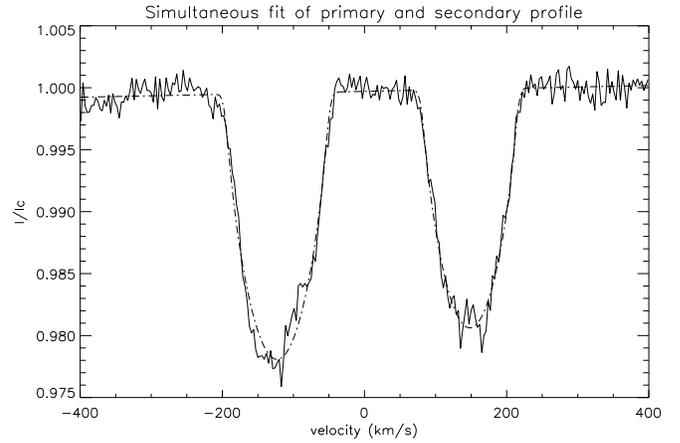}
\caption{This figure shows a binary spectrum from Jan. 12$^\mathrm{th}$ 2006 ($\Delta$ HJD = 3745.9575)  (thick line) superimposed with a simultaneous rotational fit of both components. The little bumps in the core of the line profiles are indicative of non-radial pulsations.}
\label{fig:doublefit}
\end{figure}

Fig. \ref{fig:doublefit} shows a typical LSD profile of RS Cha with both primary and secondary component well separated in radial velocity. 

\section{Orbit determination with circularized binary approximation}
\label{orbit}

\begin{figure*}[!ht]
\centering
  \includegraphics[width=13cm]{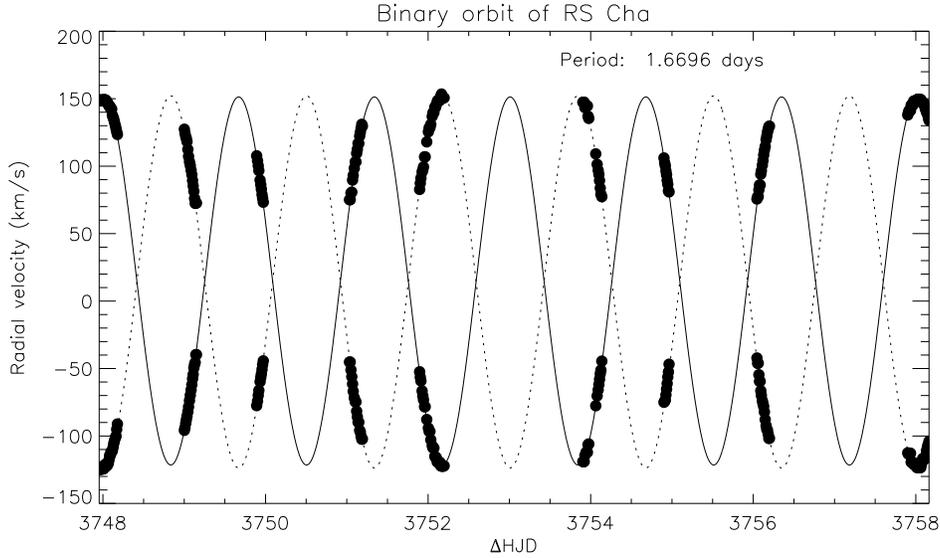}
\caption{Orbital movement of the spectroscopic binary components of RS Cha. $\Delta$HJD = HJD - 2450000.0.
The continuous and dotted lines show the adjusted circular orbits of the primary and secondary component, respectively. The dots 
represent the observed radial velocity values of both components. Dot sizes do not represent error bars and have been selected only for better visibility of the figure. Typical error bars are of the order of  50 ms$^{-1}$ and can not been represented here, due to the large scale of the orbital amplitude.}
\label{fig:orbit}
\end{figure*}

The spectroscopic binary LSD profile was then fitted with a double rotational profile (see e.g. \cite{gray}) with an IDL procedure and yielded for all cases except the fully merged profiles precise heliocentric radial velocities of the two components. In a second step, the binary orbit was then fitted with a circularized orbit (i.e. sinusoidal radial velocity curves) on the remaining 180 "binary"-LSD profiles and can be seen in Fig. \ref{fig:orbit}. Orbital parameters are listed in Table \ref{tab:orbit}. The less massive secondary component was identified based on its slightly larger orbital amplitude.

\begin{table}[!ht]
\centering
\caption{Orbital parameters of both components of RS Cha. $\Delta$HJD = HJD -  2450000.0.}
\begin{tabular}{l|c|c} \hline
Parameter & primary component & secondary component \\  \hline
$\Delta$HJD$_{\rm periastron}$    & 3750.5044 &  3749.6695   \\
Period (days)     &      1.66965$\pm$ 0.00003  &  id \\
Amplitude (km\,s$^{-1}$)    &  136.4159 & 137.9077   \\ \hline
\end{tabular}
\label{tab:orbit}
\end{table}

\begin{figure}[!ht]
  \includegraphics[width=9cm]{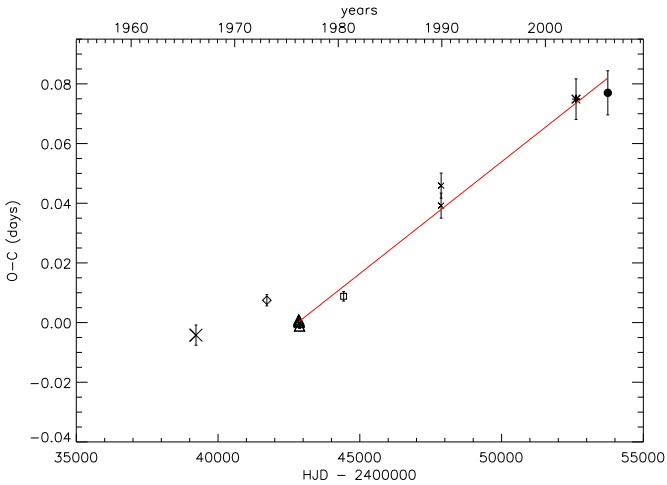}
\caption{Figure from \cite{alecian2005}. One new full circled point was added, corresponding to this new 2006 data set.}
\label{fig:omc_ad}
\end{figure}

It was interesting to determine the orbital period and its phase shift with respect to the work published previously. Based on the method described in 
Alecian et al. (\cite{alecian2005})  we calculated the observed minus computed (O-C) timings of the first conjunction and added our new point to the Figure \ref{fig:omc_ad} showing it's evolution over a time span of 30 years.  We can observe that the curve seems to decline slightly in gradient, indicative of either the presence of a third body in the RS\,Cha system or of the signature of changes in the orbital period of the order of  $\Delta P$/P = (7.5 $\pm$ 0.7) 10$^{-6}$. Further observations in the next years would be required to better characterize these orbital variations and to understand their origin. 


\section{Detection of non-radial pulsations in both components}
\label{detection}

\begin{figure}[!ht]
  \includegraphics[width=70mm]{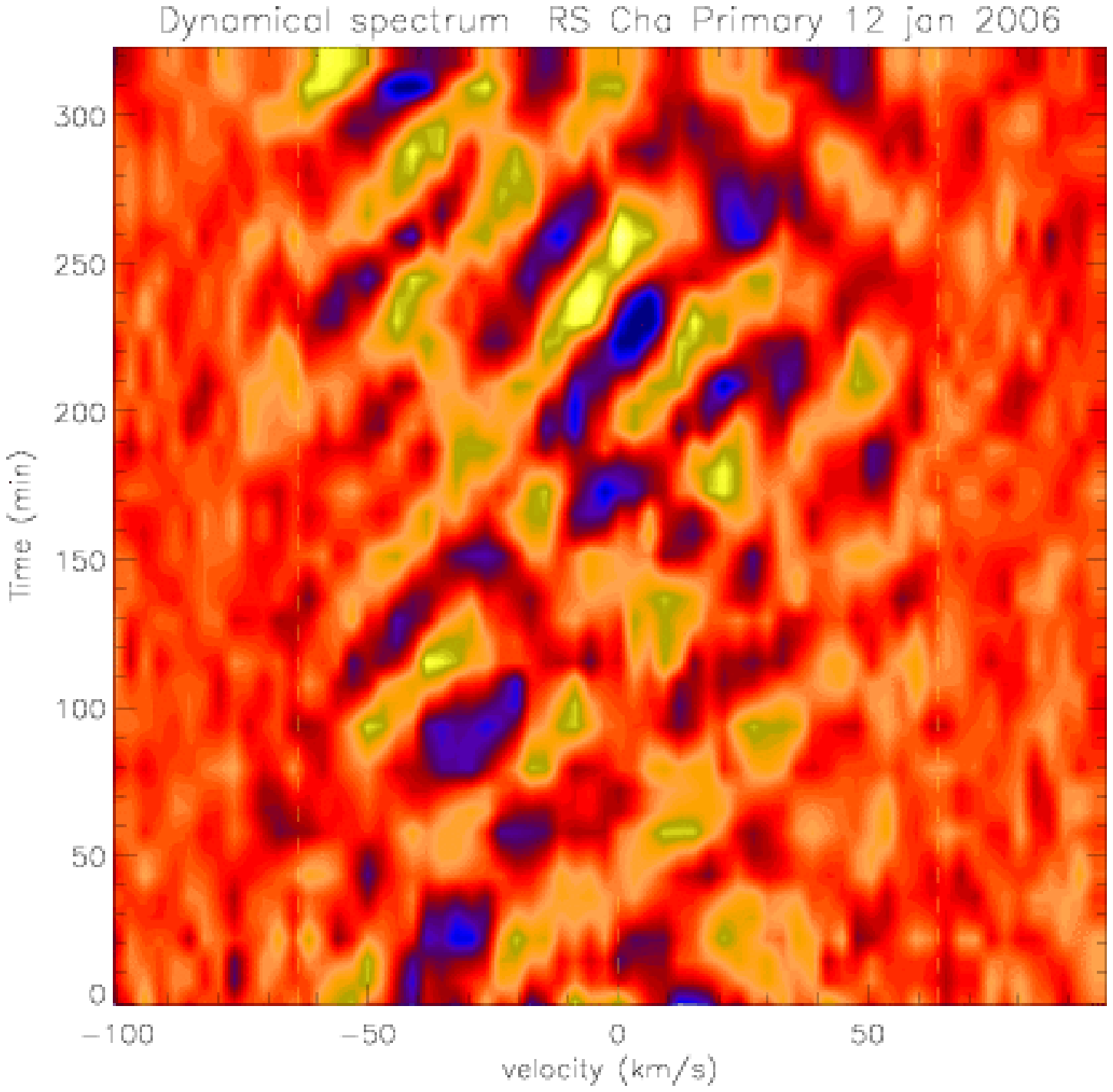}
  \includegraphics[width=70mm]{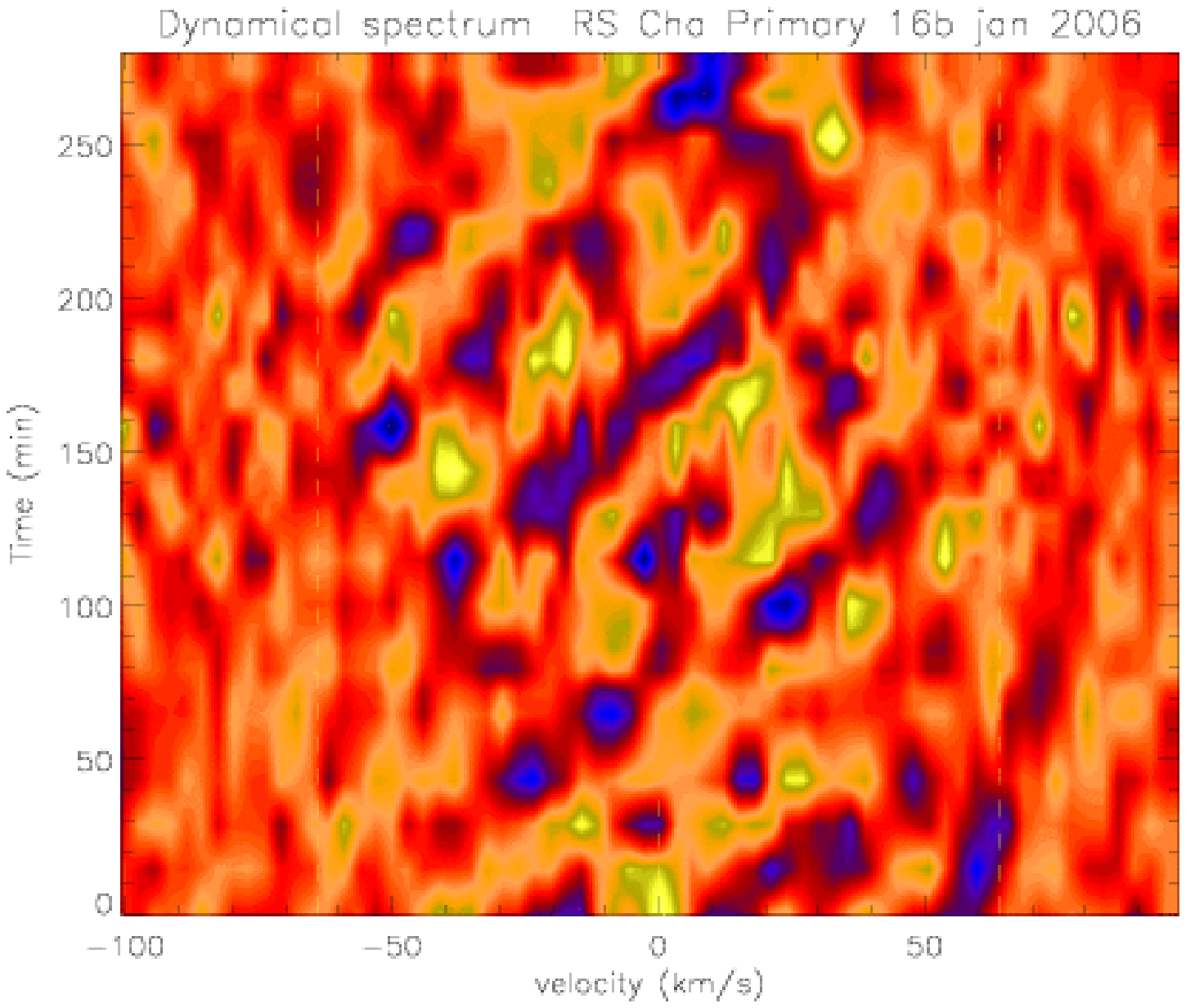}
  \includegraphics[width=70mm]{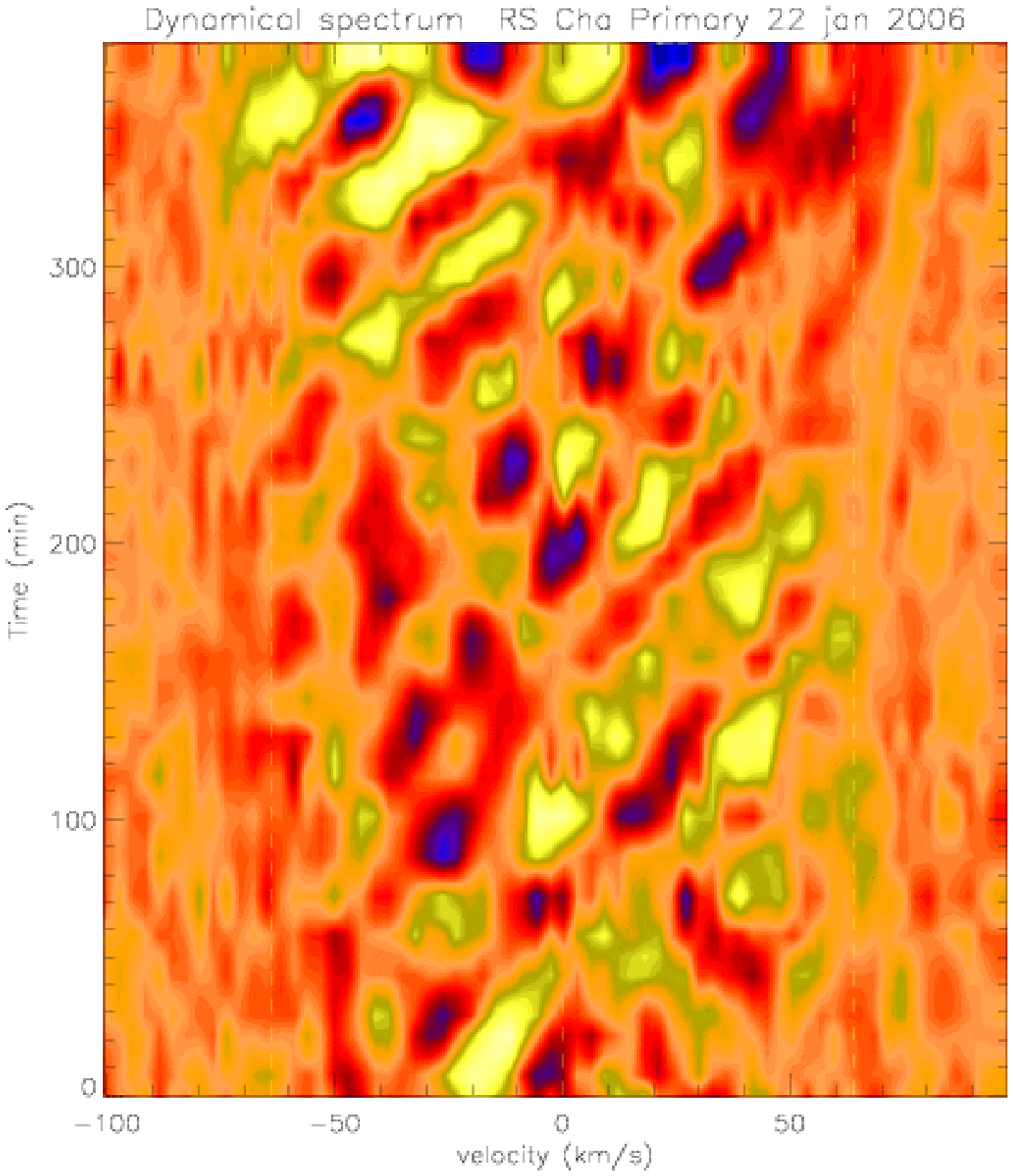}
\caption{Line profile variations due to the oscillations of RS Cha. The deviations from the mean intensity are displayed for three different nights (Jan 12$^\mathrm{th}$, 16$^\mathrm{th}$ and 22$^\mathrm{nd}$ 2006). Primary component of RS Cha.}
\label{fig:contourPrimary}
\end{figure}

\begin{figure}[!ht]
  \includegraphics[width=70mm]{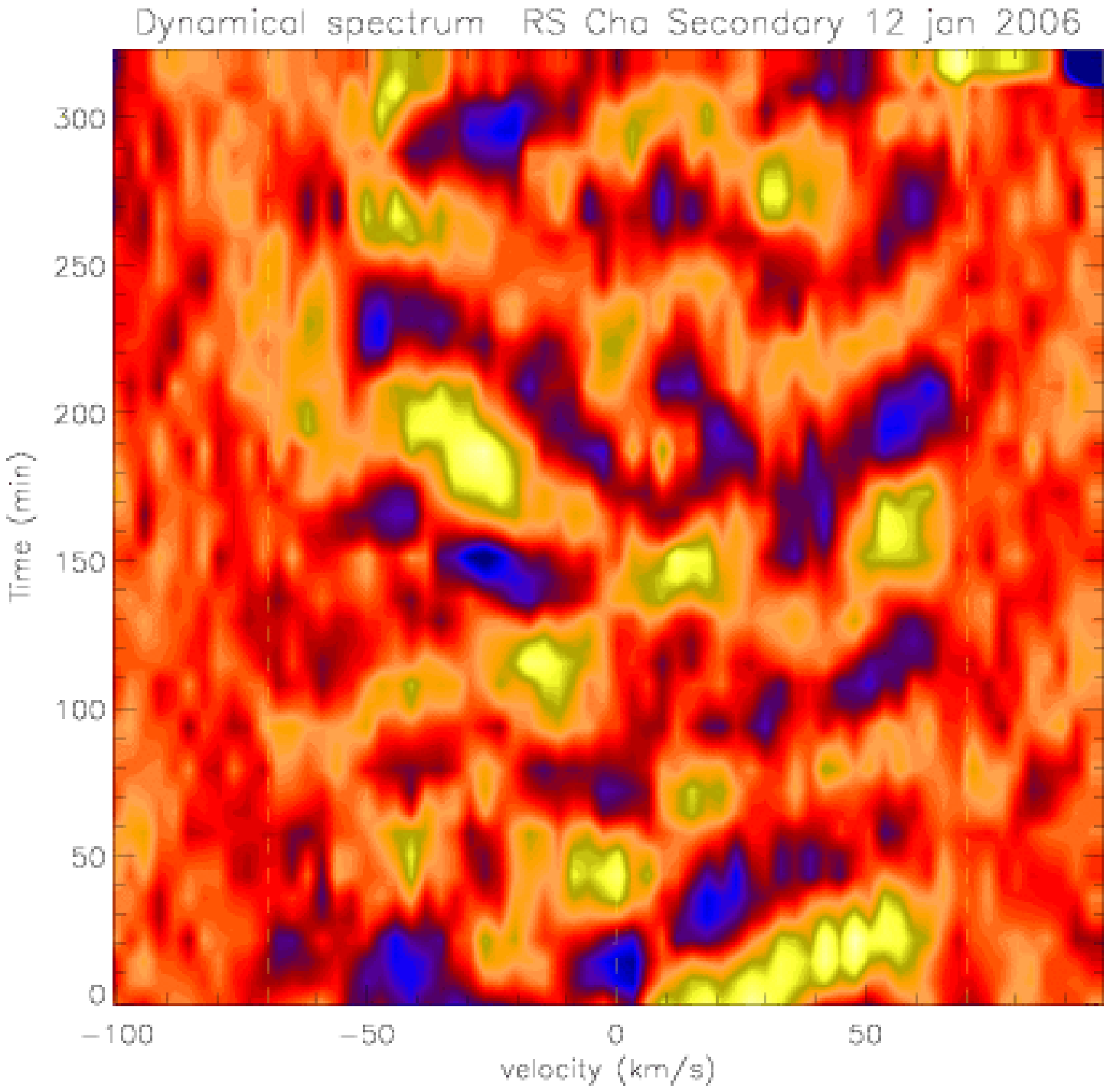}
  \includegraphics[width=70mm]{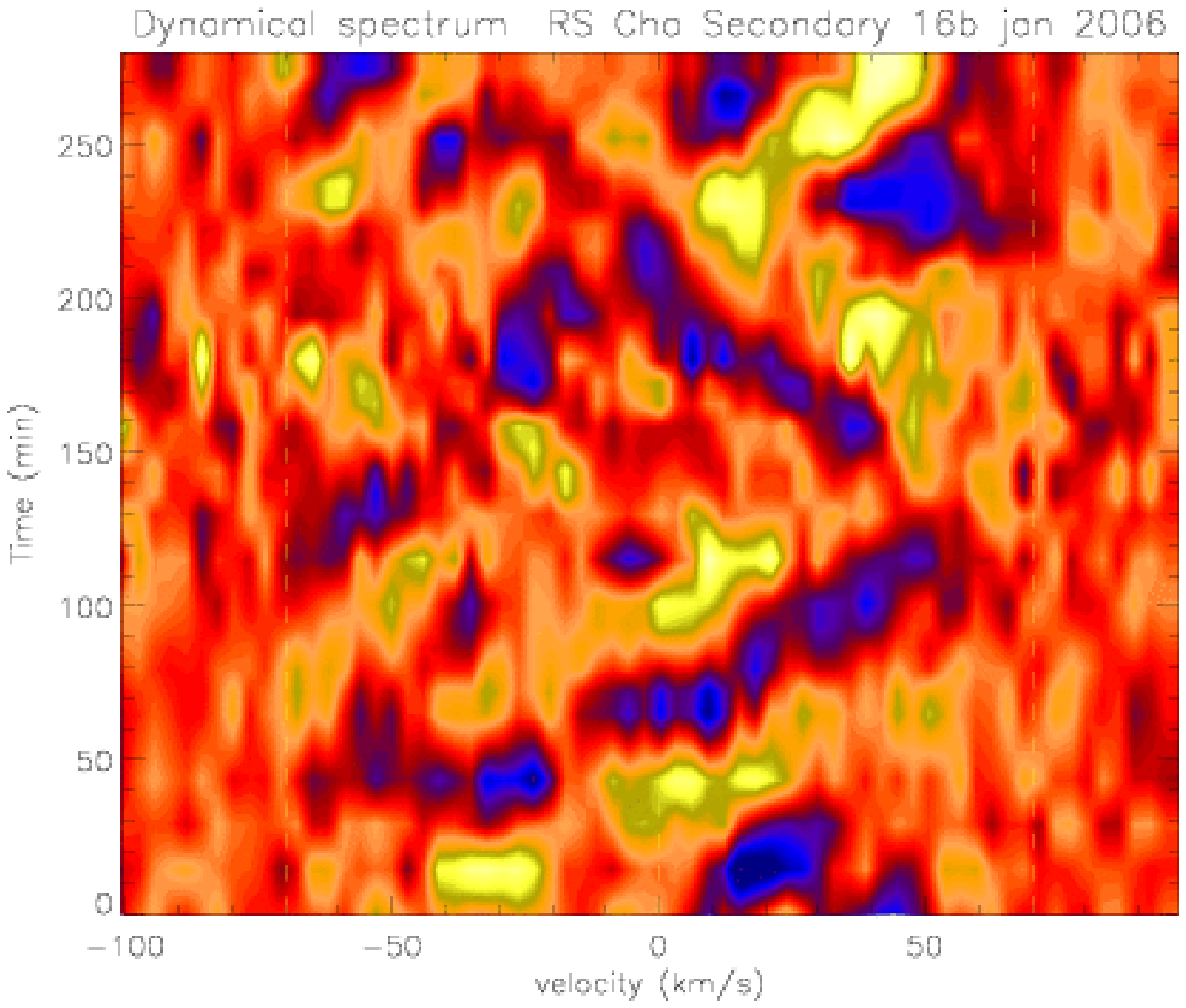}
  \includegraphics[width=70mm]{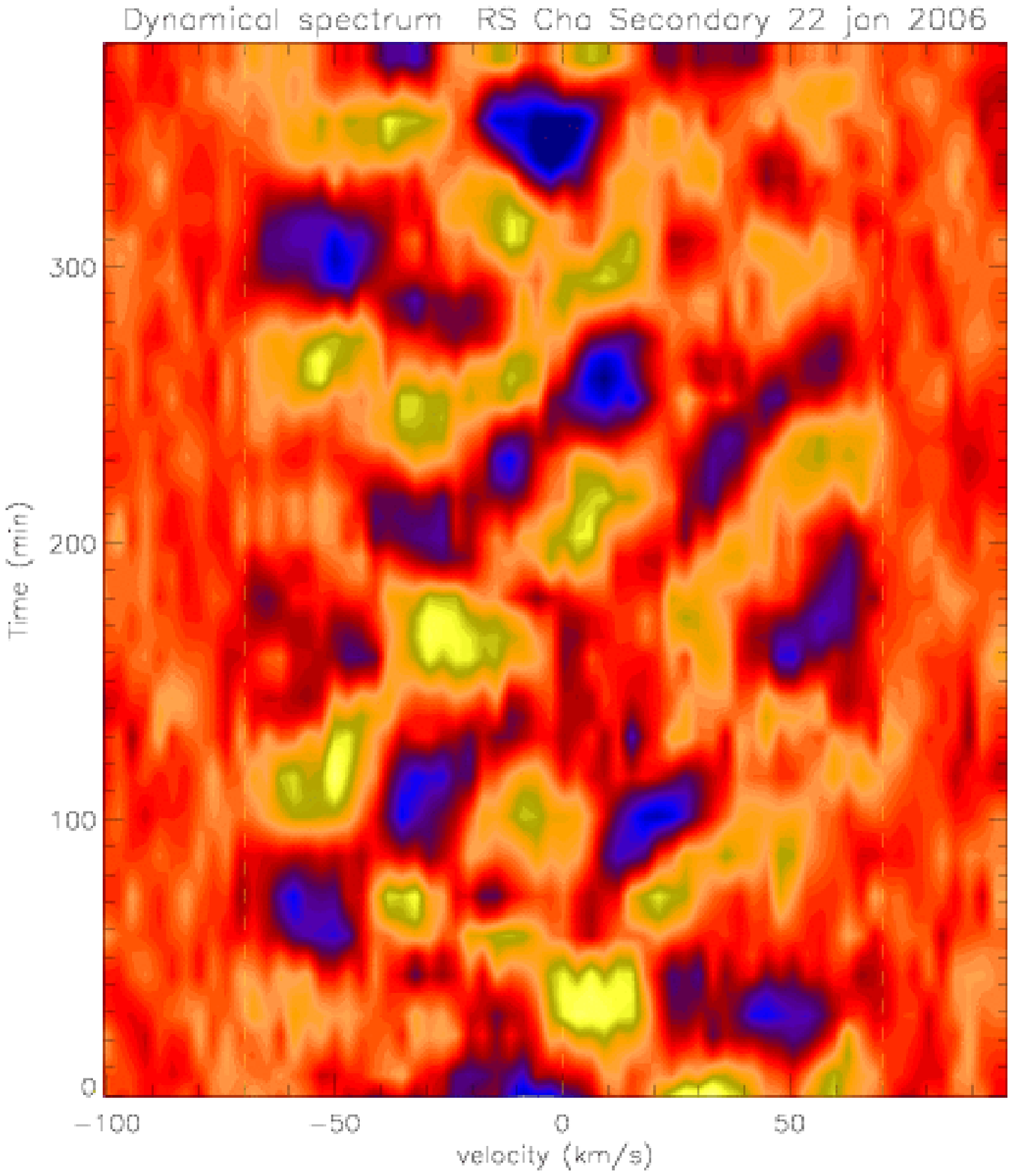}
\caption{Line profile variations due to the oscillations of RS Cha. The deviations from the mean intensity are displayed for three different nights (Jan 12$^\mathrm{th}$, 16$^\mathrm{th}$ and 22$^\mathrm{nd}$ 2006). Secondary component of RS Cha.}
\label{fig:contourSecondary}
\end{figure}

Finally, we extracted the two binary components of each LSD profile along its individual orbital radial velocity in order to obtain
two files with centred dynamical spectra revealing only the intrinsic stellar profile variations. Figs. \ref{fig:contourPrimary} and  \ref{fig:contourSecondary} reveal the deviations from the mean (rotational) profile for three different nights. It can be clearly seen that bumps move through the line profile in a quite complex manner. Fig. \ref{fig:contourPrimary} (primary) reveals a  more diagonal trend of the bump movement (bottom-left to up-right) than \ref{fig:contourSecondary} (secondary) . As will be later shown, these features are due to several high-degree non-radial pulsation modes. This represents a first direct detection of non-radial pulsations in Herbig Ae stars by spectroscopic means.

The colour coding in Figs.~\ref{fig:contourPrimary} and \ref{fig:contourSecondary} represents a dynamical range of about 0.5\% of the continuum. The first reference spectra of each night can be seen in Table~\ref{log}.

It must be noted at this stage that low frequency remnants of unproper orbital fitting subsist which can not easily be accounted for with a circularized binary orbit fit, revealing perhaps the presence of an additional yet undetected multiple component.

\section{Pulsational analysis of the primary component}
\label{pulsaprim}

A detailed search for frequencies and the subsequent identification of the pulsational modes is described in this section. At this stage, the spectra have been corrected for the radial velocity variations caused by the binary orbit of RS Cha. We selected only spectra for the frequency analysis and mode identification that do not show any blending of the LSD profiles of the two components, i.e., the situation when the difference in orbital velocities was larger than approximately the sum of the two projected rotational velocities, $\Delta v \ge v_{1} \sin i + v_{2} \sin i$. The useful Doppler velocity range for the analysis was thus between -55 and +55~\kms\ which excluded the continuum. A set of 161 spectra covering 35 hours was finally selected for the analysis. See Table~\ref{tab:selectlog} for a log of the observations.

\begin{table}
\caption[]{Finally selected data subset used for the frequency analysis. (1) 
Night  and (2) Julian date of the first observation (2,450,000+); (3) Julian 
date of the last observation (2,450,000+); (4) Number of spectra; (5) Duration 
(hrs)}                                                                           
\begin{tabular}{ccccc}                                                                                                                                       
\hline\hline                                                                                                                                                 
Date   & JD$_{\rm first}$      & JD$_{\rm last}$       & N$_{\rm spec}$        
&Duration\\                                                                  
(1)             &  (2)                          &  (3)                          
&   (4)                         &   (5)       \\                             
\hline                                                                                                                                                       
Jan 12  & 3747.9537             & 3748.1607             &  26 &  4.97\\                                                                                       
Jan 13  & 3749.0013             & 3749.1099             &  13 &  2.60\\                                                                                        
Jan 14  & 3749.8911             & 3749.9411             &  6   &  1.20\\                                                                                       
Jan 15  & 3751.0417             & 3751.1838             &  16 &  3.41\\                                                                                      
Jan 16  & 3751.8958             & 3752.1861             &  28 &  6.97\\                                                                                      
Jan 18  & 3753.9037             & 3754.1257             &  16 &  5.33\\                                                                                      
Jan 19  & 3754.9015             & 3754.9259             &  8   &  1.32\\                                                                                       
Jan 20  & 3756.0609             & 3756.1912             &  16 &  3.13\\                                                                                      
Jan 22  & 3757.9004             & 3758.1621             &  32 &  6.28\\\hline                                                                                
Total&&& 161 & 35.2\\ \hline
\end{tabular}
\label{tab:selectlog}
\end{table}

\subsection{Frequency analysis of the primary component}

We have used the software package {\sc FAMIAS} (\cite{zima2008}) for the search for periodicities in the line profile and for the identification of the pulsation modes using the Fourier parameter fit (FPF) method (\cite{zima2006}). The search for periodicities was carried out by using Discrete Fourier Transformation (DFT) and multi-periodic non-linear least-squares fitting (LSF), a method which fits a sum of sinusoidals to the variations. Both methods assume sinusoidal variations, which in general is a good approximation for pulsation modes having low amplitudes.

In order to search for pulsation frequencies \ignore{in one-dimensional time-series} we applied the following iterative procedure: in the Fourier spectrum, the frequency peak having highest amplitude is selected and the data pre-whitened with a least-squares fit that included this peak and all previously found frequencies. If several peaks have similar amplitudes, e.g., due to aliasing, the peak which results in the lowest residuals is selected. The search is stopped if none of the remaining peak reaches a S/N of above 4 (\cite{breger1993}), corresponding to a  99.9\% confidence level. 

Since the dispersion range that we selected for analysis did not extend towards the continuum, we were not able to properly compute line moments. We therefore computed the centre of gravity (CoG) of the line profile, which is numerically similar to the first moment, to study its temporal variations. Furthermore, we examined the pixel-intensity-variations (PIV) across the line profile for the frequency search. We studied the PIV by computing the mean Fourier spectrum across the line profile. The significance of a frequency peak was determined by computing its S/N in the wavelength bin where it has largest amplitude. The two methodologies are complementary in the way that the CoG is more sensitive to low-degree pulsation modes ($\ell \le 4$), whereas the PIV are also sensitive to modes of higher degree.

\begin{figure}[!ht]
\centering
\includegraphics[width=90mm,clip,angle=0]{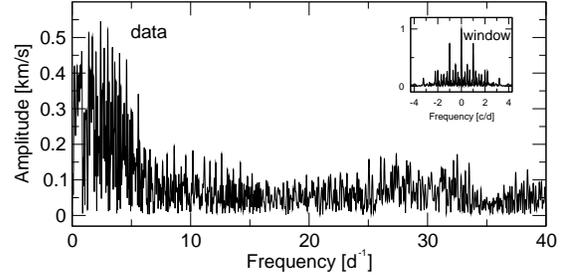}
\caption{Amplitude spectrum of the CoG time-series of the primary component. No significant frequency peaks could be detected in these data.}
\label{fig:cogprim}
\end{figure}

The Fourier spectrum of the CoG time series showed many peaks in the low-frequency domain between 0 and 6~\cd (see Figure~\ref{fig:cogprim}). These peaks are associated to nightly zero point velocity shifts of the line profile and are likely caused by the uncertainty of the binary orbit fitting. No significant peaks were found above 6~\cd.

\begin{figure}[!ht]
\centering
\includegraphics[width=90mm,clip,angle=0]{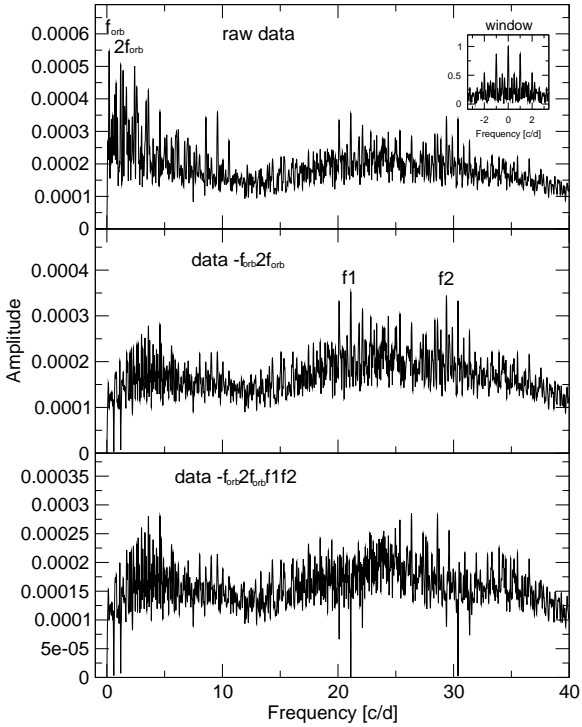}
\caption{Amplitude spectrum of the pixel-by-pixel intensity time-series of the primary component. The different panels show the data after subsequent pre-whitening of the indicated frequency peaks.}
\label{fig:pbpprim}
\end{figure}

The frequency analysis of the PIV also revealed several peaks in the low frequency domain (see Figure~\ref{fig:pbpprim}). Here, they are clearly associated with the uncertainties of the orbit fitting. We found significant frequencies at 0.599 and at 1.197~\cd, which corresponds to the orbital frequency and twice thereof. After pre-whitening with these terms, two other significant frequencies could be detected in the data: $f_1=21.11$ and $f_2=30.38$~\cd. After further pre-whitening, no additional frequencies having a S/N above 4 could be found. Table~\ref{tab:freqprim} lists the frequencies detected in the primary component.

\begin{table}[!ht]
\begin{center}
\caption{Frequencies detected in the pixel-by-pixel intensity variations of the primary component of RS Cha. The value $A_\mathrm{PIV}$ is the integral of the pixel-amplitude across the line profile between -55 and 55 \kms. For each frequency, the S/N is computed for the wavelength bin where its amplitude is highest.}
\begin{tabular}{l|c|c|c} \hline
ID      & freq.    & $A_\mathrm{PIV}$ & S/N\\
        & \cd      & \kms &\\ \hline
$f_{\mathrm{orb}}$    &    0.599 & 0.106 & 13.9 \\
$2f_{\mathrm{orb}}$   &    1.197 & 0.040 & 4.5 \\
$f_1$                 &    21.11 & 0.040 & 4.5 \\  
$f_2$                 &    30.38 & 0.038 & 4.0 \\\hline 
\end{tabular}
\end{center}
\label{tab:freqprim}
\end{table}

\subsection{Mode identification with the FPF method}
We carried out a mode identification of the two detected frequencies with the FPF method. This method makes use of the fact that stellar oscillations cause intensity variations in a rotationally broadened line profile due to the Doppler effect. It is capable to determine the spherical degree, $\ell$, and the azimuthal order, $m$, of a pulsation mode from a time series of line profile variations. The contribution of each pulsation frequency to the observed line profile variations can be de composed into its Fourier parameters, the zero-point, amplitude, phase across the line profile. These Fourier parameters are determined from a multi-periodic least-squares fit across the line profile, i.e., for each wavelength bin, and have a characteristic distribution according to the related $\ell$ and $m$-values. By comparing the observed Fourier parameters and their counterparts computed from synthetic line profile variations, we can determine the pulsational characteristics of a pulsation mode. The reduced chi-square values $\chi^2_\nu$ of the different solutions are computed from the complex amplitudes and their uncertainties across the line profile. For details we refer to \cite{zima2006}.

The synthetic line profiles were computed assuming that the pulsations can be described by a superposition of spherical harmonics. Furthermore, we assume a Gaussian intrinsic line profile, one infinitesimal thin line forming region, and a symmetry axis of the pulsation field that coincides with the rotation axis.

We searched the following parameter space [min; max; step]: $\ell\in[0; 15; 1]$, $m\in[-\ell; \ell; 1]$ for each $\ell$, pulsation velocity amplitude $a\in[0; 60; 0.5]$~\kms, \vsini$\in[30; 80; 1]$, and width of Gaussian intrinsic profile $\sigma\in[1; 15; 1]$~\kms. The inclination value of the RS Cha system is well known (\cite{clausen}). Assuming that the inclination of the rotation axis of the components is identical to that of the system, we set it as fixed parameter to $i=83.4^\circ$ during the search for the optimum in the parameter space.

\begin{figure*}[!ht]
\centering
\begin{center}
 \begin{tabular}{cc}
$f_1=21.11$~\cd & $f_2=30.38$~\cd\\
  \includegraphics*[width=60mm,clip,angle=0]{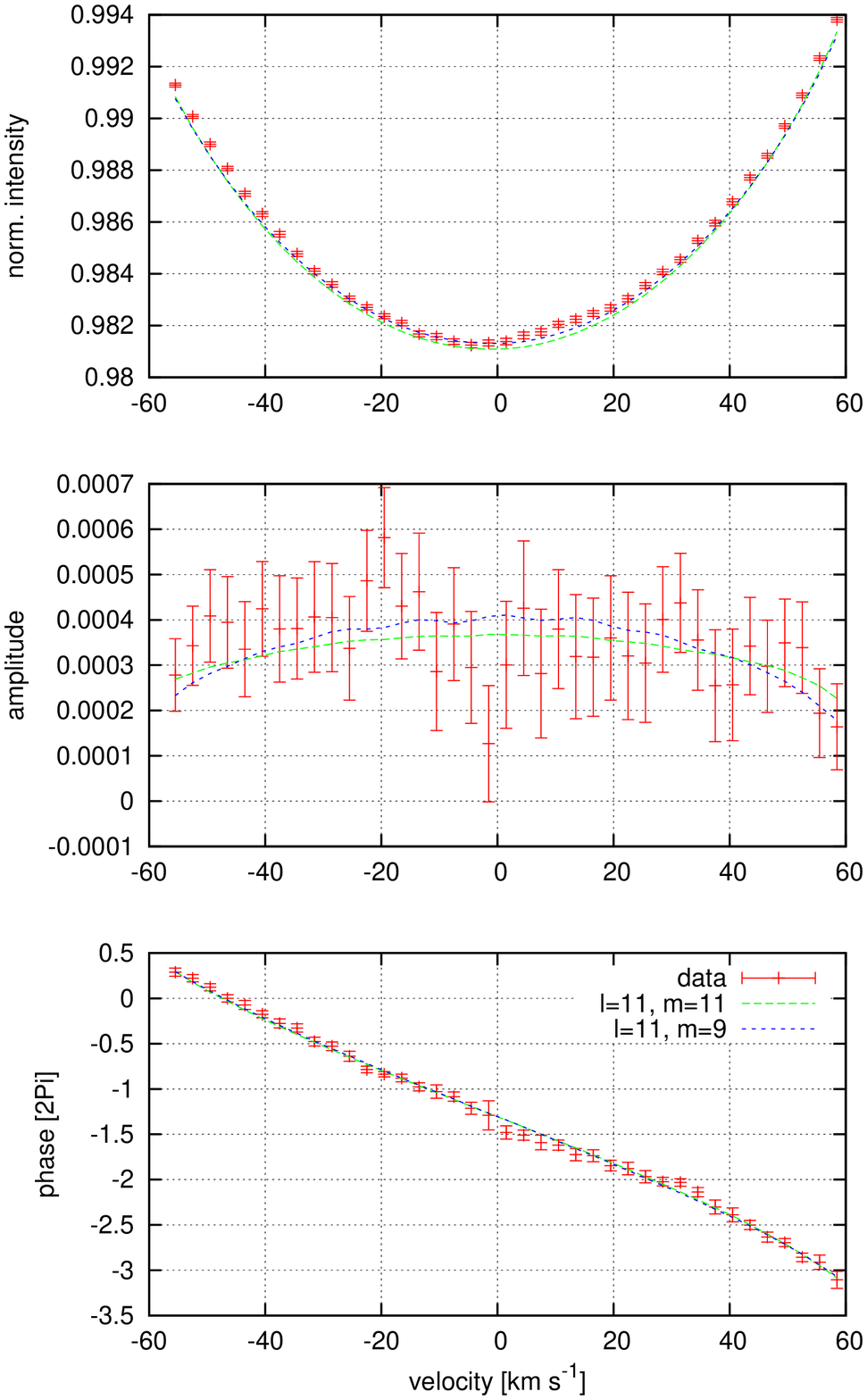}&
  \includegraphics*[width=60mm,clip,angle=0]{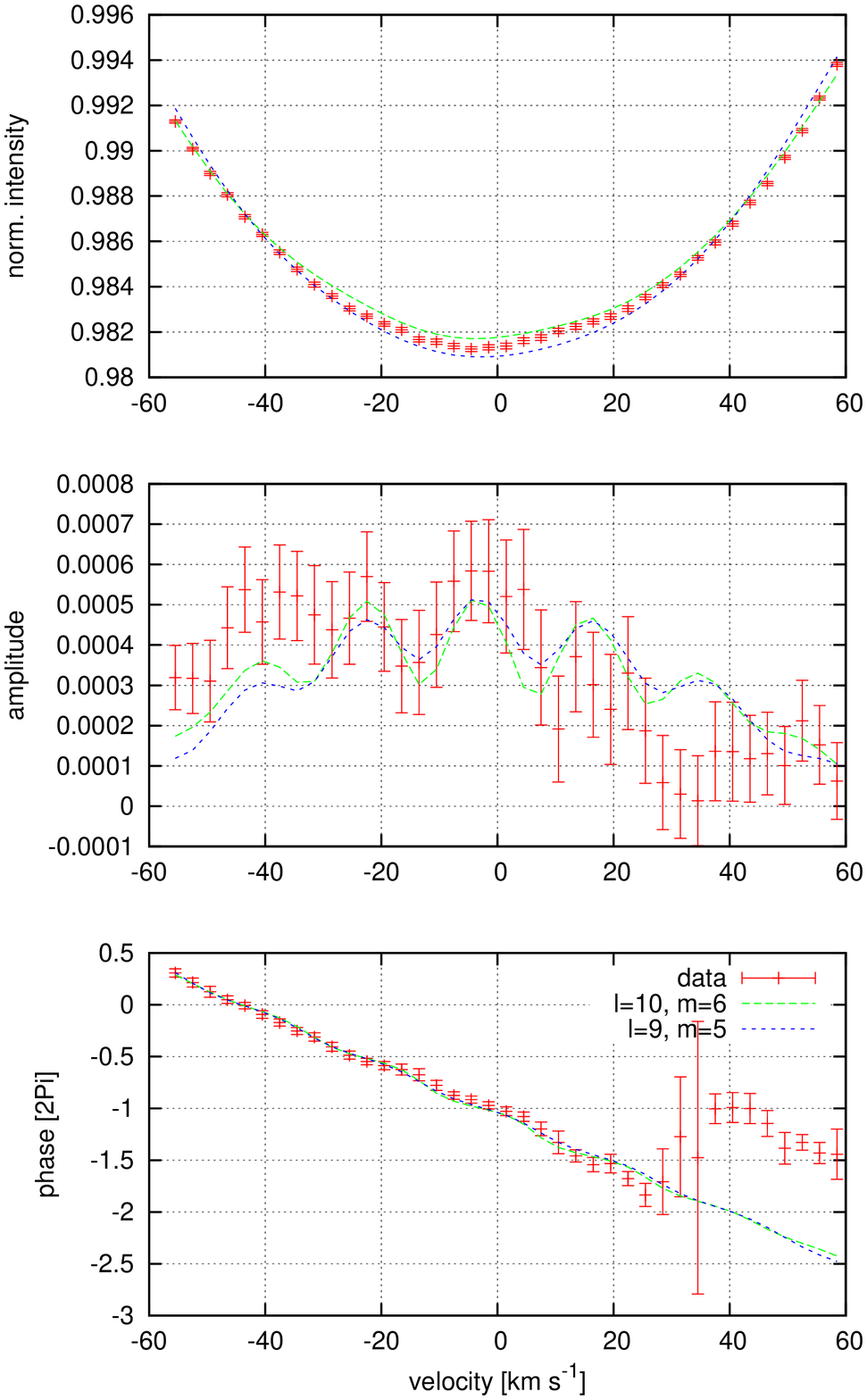}\\
\end{tabular}
\end{center}
\caption{Fourier parameters across the line profile of the detected pulsation frequencies. Each panel shows from top to bottom: zero-point, amplitude and phase. For both frequencies, the two best fits to the observed zero point, amplitude in units of the continuum and the phase distribution in units of $2\pi$ are shown. The uncertainties are the formal errors computed from the least-squares fitting. Primary component of RS Cha.}
\label{fig:fitzap1}
\end{figure*}

Our analysis with the FPF method revealed that both detected frequencies are high-degree pro-grade pulsation modes. Due to the uncertainties of the derived amplitude and phase across the line profile, we were not able to unambiguously identify $\ell$ and $m$ but are left with a number of solutions, which still will constrain possible theoretical seismic models. The five best solutions of each frequency are listed in Table~\ref{tab:fpfprimary}. Figure~\ref{fig:fitzap1} shows the Fourier parameters of the observed and best fitting model to the two detected frequencies.

We identified both frequencies as pro-grade modes of high degree. The best solution for $f_1$ identifies it as sectoral mode of $\ell=m=11$, whereas the best identification for $f_2$ is a tesseral mode with $\ell=10$ and $m=6$. The uncertainties of this identification are larger for $f_2$ than for $f_1$ and in the range of $\pm 1$ for both quantum numbers.

We derived a projected rotational velocity of \vsini=$68 \pm 2$~\kms\ which agrees quite well with the previously determined value by Alecian et al. (2005) and implies a rotation frequency of $0.63 \pm 0.01$~\cd. 

\begin{table}[!ht]
\centering
\normalsize
\caption{Mode parameters derived from the application of the FPF method. For each pulsation mode, the four best solutions are shown. $\chi^2_\nu$ is the reduced chi-square value, $a$ is the velocity amplitude in~\kms;  $\sigma$ is the width of the intrinsic Gaussian profile, both expressed in \kms. The inclination value was fixed at 83.4$^\circ$; Primary component of RS Cha.}

\begin{tabular}{cccccc}
\hline
$\chi^2_\nu$ & $\ell$ & $m$ & $a$  &  \vsini & $\sigma$ \\  \hline
\multicolumn{6}{c}{$f_1=21.11$~\cd} \\ \hline
0.70 & 11 & 11 & 0.5 & 68 & 3\\
0.74 & 11 & 9  & 1.5 & 68 & 4\\
0.79 & 10 & 6  & 4.0 & 67 & 3\\
0.85 & 11 & 7  & 4.5 & 71 & 5\\
1.04 & 10 & 8  & 1.5 & 66 & 5\\
\hline\hline
\multicolumn{6}{c}{$f_2=30.38$~\cd} \\ \hline
1.21 & 10 & 6 & 52.0  & 67 & 12\\
1.33 & 9  & 5 & 20.0  & 66 & 11\\
1.37 & 11 & 7 & 50.0  & 71 & 12\\
1.41 & 8  & 4 & 9.5   & 66 & 7\\
1.48 & 11 & 9 & 43.0  & 70 & 13\\
\hline\hline

\hline
\end{tabular}
\label{tab:fpfprimary}
\vspace*{5mm}
\end{table}

\subsection{Mode identification with the Fourier 2D method}

A complementary, and very direct method of analysing the non radial pulsation modes present in the LSD-spectra time series consists of applying the Fourier 2D method (\cite{kennelly}, \cite{kennelly2}), resulting in a representation of the data in the frequency - apparent $m$ space. The apparent $m$ is related to the structure of the modes present at the stellar surface, without being identical to the usual azimuthal order m. However, the original work by \cite{kennelly} showed that apparent  $\left | m \right |$ scales as $\ell$ +2 for values close to zero, as $\ell$ +1 for values lower than 10 and as $\ell$ for values above 10. Basically, this method consists of doing subsequently a spectral and a temporal Fourier analysis. All LSD profiles of the time series were normalized to the same equivalent width, in order to eliminate the effects of large scale variations not related to the signature of pulsations. The time average of the mean profiles was then constructed and subtracted from each mean profile, yielding a time series of  normalized residual profiles. Before doing the 2 dimensional Fourier analysis, each residual profile was interpolated on a grid representing stellar longitudes, transforming velocities across the line profile into longitudes on the stellar equator using the relation  $\Delta v = v\sin i  \sin\phi$, where 
$\Delta v$ is the velocity position within the LSD profile with respect to the rest wavelength of the star and $\phi$ the stellar longitude angle of the star in spherical coordinates (under assumption that $\sin i$ is close to unity, which is the case for RS Cha). The Fourier transform being very sensitive to gaps in the temporal sampling, we selected only continuous series of spectra within a night, and interpolated them on a $2^{n}$ regularly spaced temporal grid, where n is selected in order to avoid over- or under-sampling. As a result, we obtained F2D spectra for each suitable night. In order to take into account the individual quality of each night, we decided to combine the spectra of the three nights with a time-basis of more than 250 continuously observed minutes in the following way:

\begin{equation}
F2D_{\mathrm{tot}} = \sqrt{\displaystyle\sum_{j=1}^{n}  a_j    \left |{ F2D_{j}}\right |^{2}   }     
\end{equation}

where $a_j$ is a normalized weighting coefficient:

\begin{equation}
a_j =               \frac{n_{\mathrm{spec}_j} \overline {S/N}_j^2}{\displaystyle\sum_{k=1}^{n}   n_{\mathrm{spec}_k}\overline {S/N}_k^2 }
\end{equation}

and      $\overline{S/N}_j$ the average signal to noise ratio per spectrum for a given night, $n_{\rm spec}$ the number of spectra per night and n the number of exploitable nights (in the sense of the Fourier 2D analysis). The coefficient a$_j$ reflects therefore the total cumulated stellar photon flux per night, normalized to the cumulated photon flux of the full observing run. While this expression is not mathematically  rigorous it reflects to some extent the quality of the individual nights. 

\begin{figure}[!ht]
  \includegraphics[width=70mm]{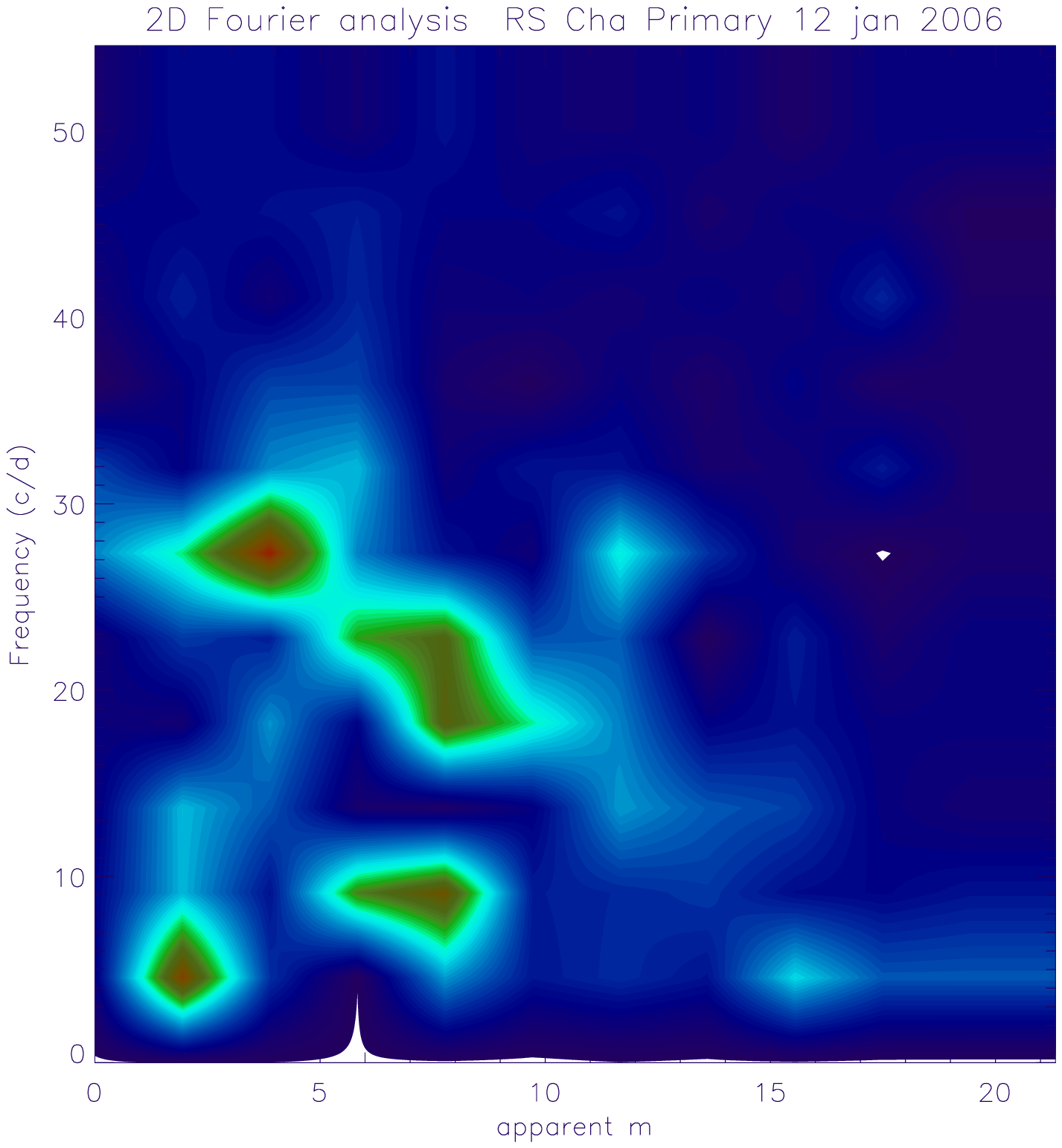}
  \includegraphics[width=70mm]{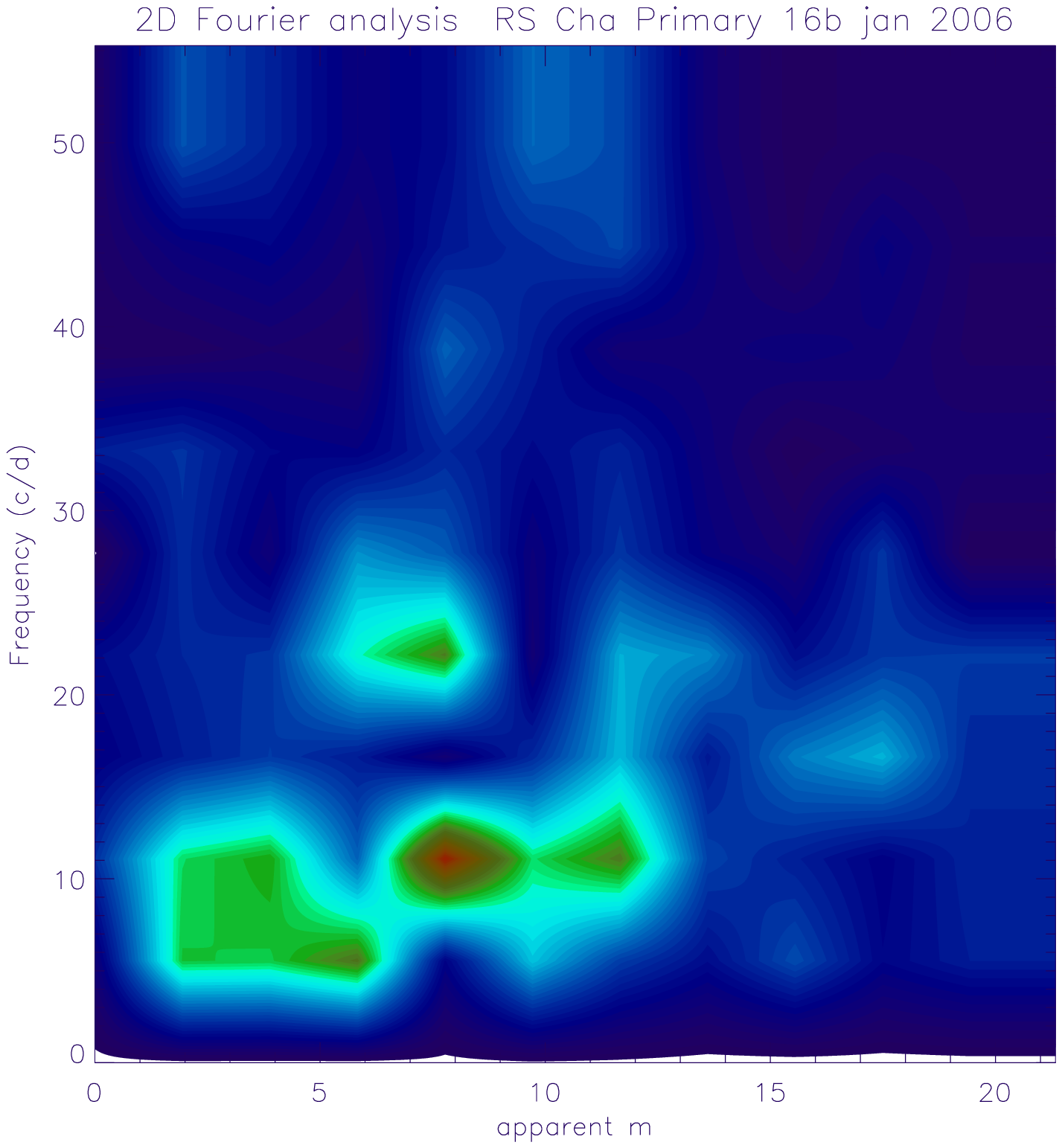}
  \includegraphics[width=70mm]{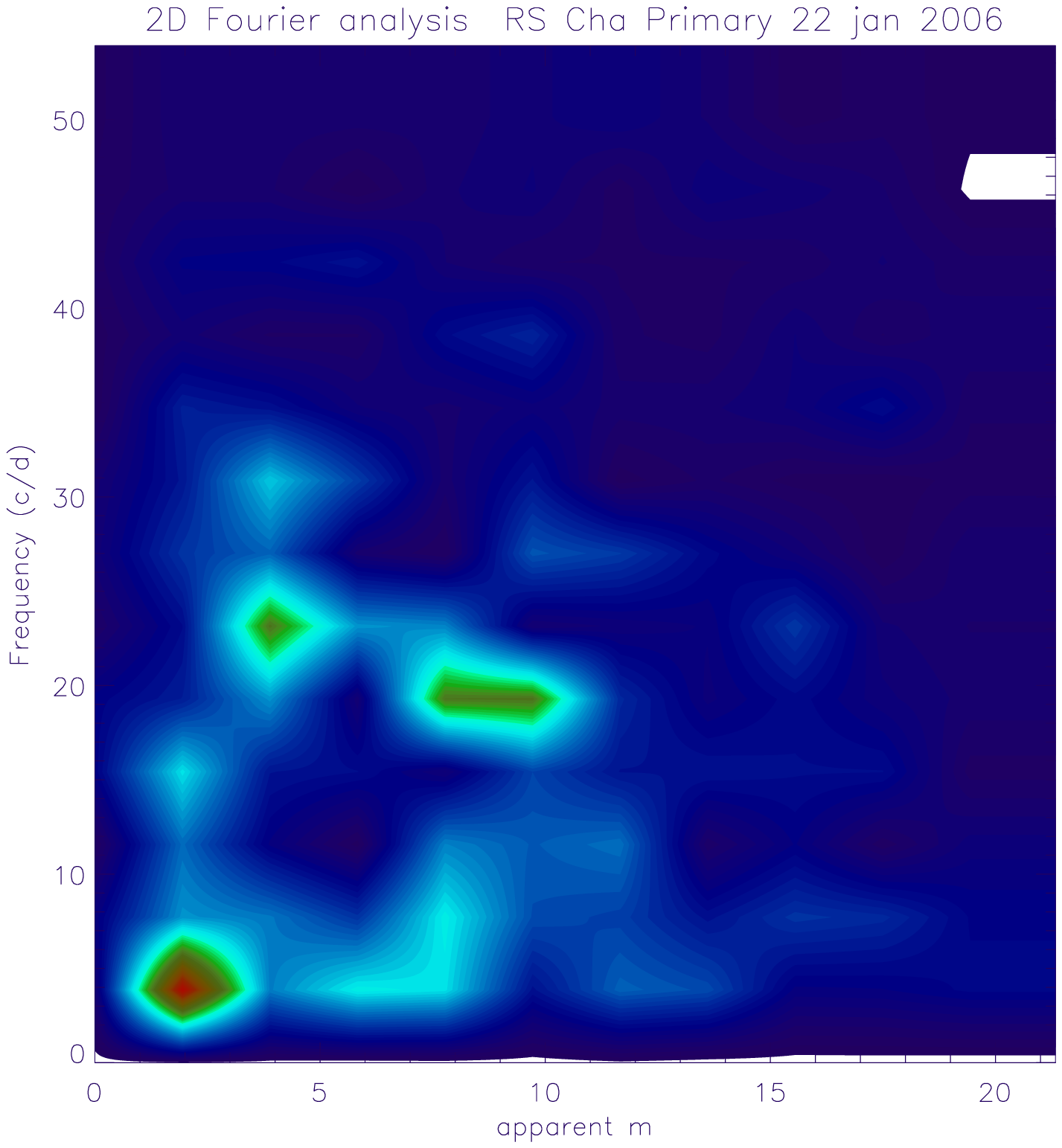}
\caption{F2D analysis on the Primary component of RS Cha for the nights of Jan 12$^\mathrm{th}$, 16$^\mathrm{th}$ and 22$^\mathrm{nd}$ 2006.  Apparent $\left | m \right |$ scales as $\ell$ +2 for values close to zero, as $\ell$ +1 for values lower than 10 and as $\ell$ for higher values (\cite{kennelly}). }
\label{fig:f2D_Primary}
\end{figure}

\begin{figure}[!ht]
  \includegraphics[width=70mm]{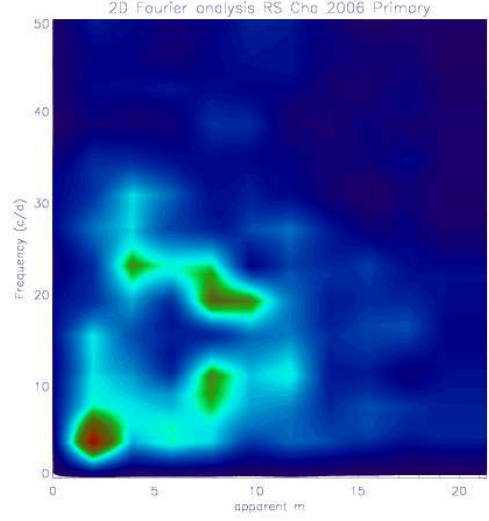}
\caption{Weighted F2D periodogram of the Primary component of RS Cha for the nights of Jan 12$^\mathrm{th}$, 16$^\mathrm{th}$ and 22$^\mathrm{nd}$ 2006.}
\label{fig:f2D_Primary_weighted}
\end{figure}

Figure~\ref{fig:f2D_Primary} shows the nightly F2D analysis for the nights of 12$^\mathrm{th}$, 16$^\mathrm{th}$ and 22$^\mathrm{nd}$ of January 2006. \cite{kumu} estimate the uncertainty in the frequency to be approximately 1/(4$\Delta T$), where $\Delta T$ is the time span of the data set. In our case for a typical observing night of 300\,min continuous observations,   this would yield an uncertainty of 1.2\,d$^{-1}$ in the frequency determination. This, together with large-scale beating effects on the pulsational signature and significant S/N ratio variations from one night to another might account for slightly different positions of frequencies in the F2D space. However, the "general" pattern is preserved from night to night, as can be seen in Figure~\ref{fig:f2D_Primary}, and the weighted F2D periodogram in Figure ~\ref{fig:f2D_Primary_weighted} 
exhibits the strongest recurring oscillation modes.

Table \ref{tab:fou2dprimary} summarizes the results for the primary component based on the F2D analysis.

\begin{table}
\caption[]{Frequency search and mode identification. Summary of the weighted F2D analysis on the primary component. 
Columns are as following: (1) frequency (error bars are of the order of 1.2\,d$^{-1}$) (2) possible identification with FPF result, (3) apparent $\left | m \right |$, (4) expected $\ell$ value.\\}
\begin{tabular}{cccc}
\hline\hline
       d$^{-1}$        &  FPF frequency & app.  $\left | m \right |$ & $\ell$\\\hline
       20.		  &     	$f_1$     &       	8-10  &  8-10 \\
       24.		  &		               &		4	 &   2 or 3      \\
       12.		  &		               &          8	 &   8   \\
       4.		  &			      &          2        &   0  \\ \hline
\end{tabular}
\label{tab:fou2dprimary}
\end{table}

The F2D method confirms the existence of the high degree mode $f_1$ determined by the FPF method ($f_1$ = 21.11\,d$^{-1}$) and indicates a degree $\ell$ of 8-10, within the error bars of the FPF determination. While single night F2D periodograms show some high frequency signal around 29.\,d$^{-1}$ (e.g. night of Jan 12$^\mathrm{th}$), which could correspond to $f_2$ (FPF), this is not confirmed systematically. A low frequency signal around 4.\,d$^{-1}$ is very likely a fundamental oscillation frequency due to the window function (t$_{obs}$ typically 1/4 d) and is present in all F2D periodograms seen in Figure~\ref{fig:f2D_Primary}. Other frequencies found in the weighted F2D periodogram around 12.\,d$^{-1}$ and 24.\,d$^{-1}$ (with different apparent $\left | m \right |$ values) have not been revealed by the FPF method.

\section{Pulsational analysis of the secondary component}
\label{pulsasecond}

The analysis of the secondary component was performed on a selected data subset, following the same selection rules as for the primary component. This procedure left 165 spectra that were suited for the further analysis. Also for the second component,  we had to cut out the central parts of the line profile between $\pm 70$~\kms\ in Doppler velocity. Therefore, we also had to use the CoG of the profile to estimate the radial velocity variations. We carried out the frequency search and mode identification in the same manner as for the primary component.

\subsection{Frequency analysis of the secondary component}

\begin{figure}[!ht]
\centering
\includegraphics[width=90mm,clip,angle=0]{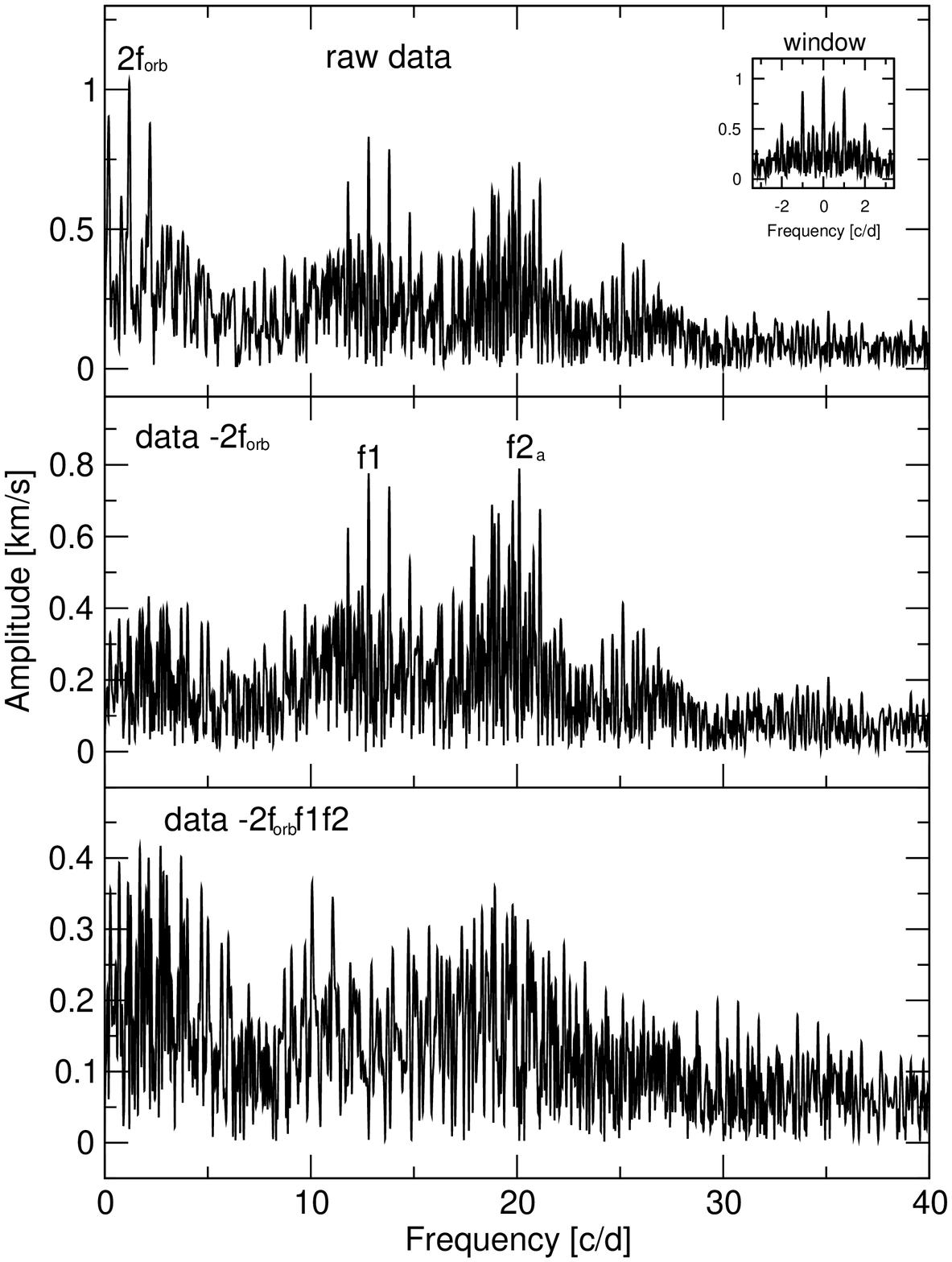}
\caption{Amplitude spectrum of the CoG time-series of the secondary component. The different panels show the data after subsequent pre-whitening of the indicated frequency peaks.}
\label{fig:cogsec}
\end{figure}

The Fourier spectrum of the CoG time series (see Figure~\ref{fig:cogsec}) shows a dominant frequency peak with an amplitude of 1~\kms\ at twice the orbital frequency $1.19$~\cd. After pre-whitening of this frequency, the low frequency domain below 6~\cd\ does not show any more significant frequencies and has a much lower noise level than the primary component. Strong 1~\cd-aliasing complicates the detection of further frequencies. The highest remaining peak is at 12.8~\cd\ with an amplitude of 0.78~\kms, but the one-day aliasing peak at 13.9~\cd\ has a similar amplitude. A similar situation occurs at 20.1~\cd. We selected two peaks that had a S/N above 4 and that resulted in the lowest residuals after pre-whitening, $f_1=12.81$ and $f_{\mathrm 2a}=20.11$~\cd. After pre-whitening with these frequencies, no further significant peaks can be found in the Fourier spectrum.

\begin{table}[!ht]
\begin{center}
\caption{Frequencies detected in the pixel-by-pixel intensity variations of the secondary component of RS Cha. The value $A_\mathrm{PIV}$ is the integral of the pixel-amplitude across the line profile between -70 and 70 \kms. For each frequency, the S/N is computed for the wavelength bin where its amplitude is highest.}
\begin{tabular}{l|c|c|c} \hline
ID      & freq.    & $A_\mathrm{PIV}$ & S/N\\
        & \cd      & \kms &\\ \hline
$f_{\mathrm{orb}}$    &    0.599  & 0.060 & 5.7 \\
$2f_{\mathrm{orb}}$   &    1.197  & 0.138 & 8.2 \\
$f_1$                 &    12.81  & 0.066 & 3.9\footnote{This frequency has a S/N=5 in the CoG time-series. Therefore, we relaxed the significance criterion for this frequency in the PIV.} \\  
$f_{\mathrm 2b}$                &    19.11  & 0.061 & 5.7 \\
$f_3$                 &    24.56  & 0.054 & 4.0 \\\hline 
\end{tabular}
\end{center}
\label{tab:freqsec}
\end{table}

\begin{figure}[!ht]
\centering
\includegraphics[width=90mm,clip,angle=0]{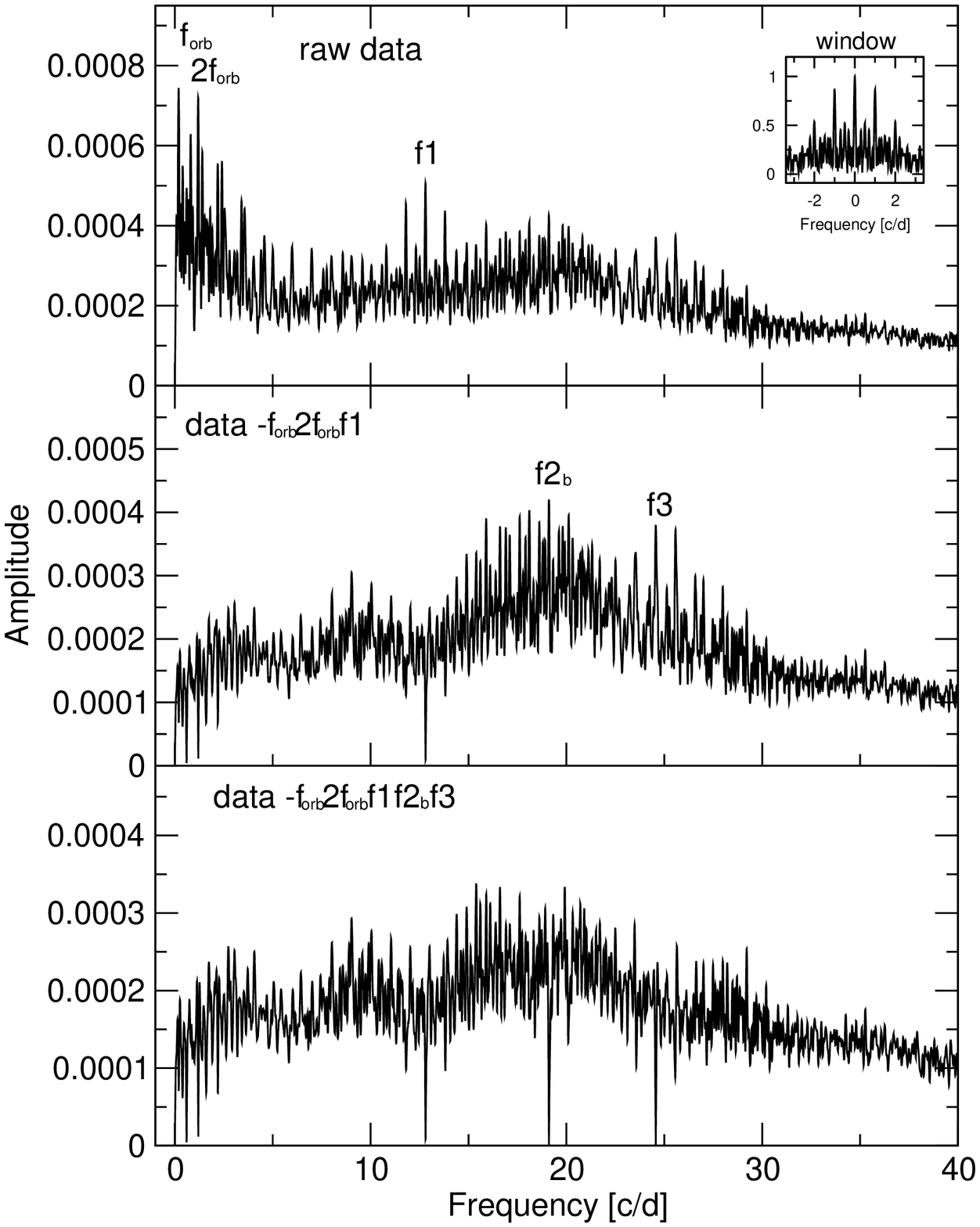}
\caption{Amplitude spectrum of the pixel-by-pixel intensity time-series of the secondary component. The different panels show the data after subsequent pre-whitening of the indicated frequency peaks.}
\label{fig:pbpsec}
\end{figure}

The analysis of the PIV across the line profile (see Figure~\ref{fig:pbpsec}) revealed a dominant peak at 0.19~\cd, which is a one-day aliasing of twice the orbital frequency at 1.19~\cd. 
Also the orbital frequency had a significant amplitude in the PIV which, as has been mentioned before, is a result of imperfect orbital fitting. We pre-whitened the data with these two frequencies and found $f_1=12.81$~\cd\ as highest remaining peak. After further pre-whitening, the highest peak is at $f_{\mathrm 2b}=19.11$~\cd\ which is a one-day aliasing of $f_{\mathrm 2a}$. Pre-whitening with $f_{\mathrm 2b}$ resulted in much lower residuals than pre-whitening with $f_{\mathrm 2a}$, which contradicts the frequency analysis of the CoG. Since we used the PIV for the mode identification with the FPF method, we accepted $f_{\mathrm 2b}$ as the 'true' frequency. After pre-whitening, one more frequency at $f_3=24.56$~\cd\ with a S/N of 4 was detected. No other significant peaks could be extracted from the data. The detected frequencies are listed in Table~\ref{tab:freqsec}.

\subsection{Mode identification with the FPF method}

\begin{figure*}[!ht]
\centering
\begin{center}
 \begin{tabular}{ccc}
$f_1=12.81$~\cd & $f_{\mathrm 2b}=19.11$~\cd & $f_3=24.56$~\cd\\
  \includegraphics*[width=55mm,bb=50 50 490 764,clip,angle=0]{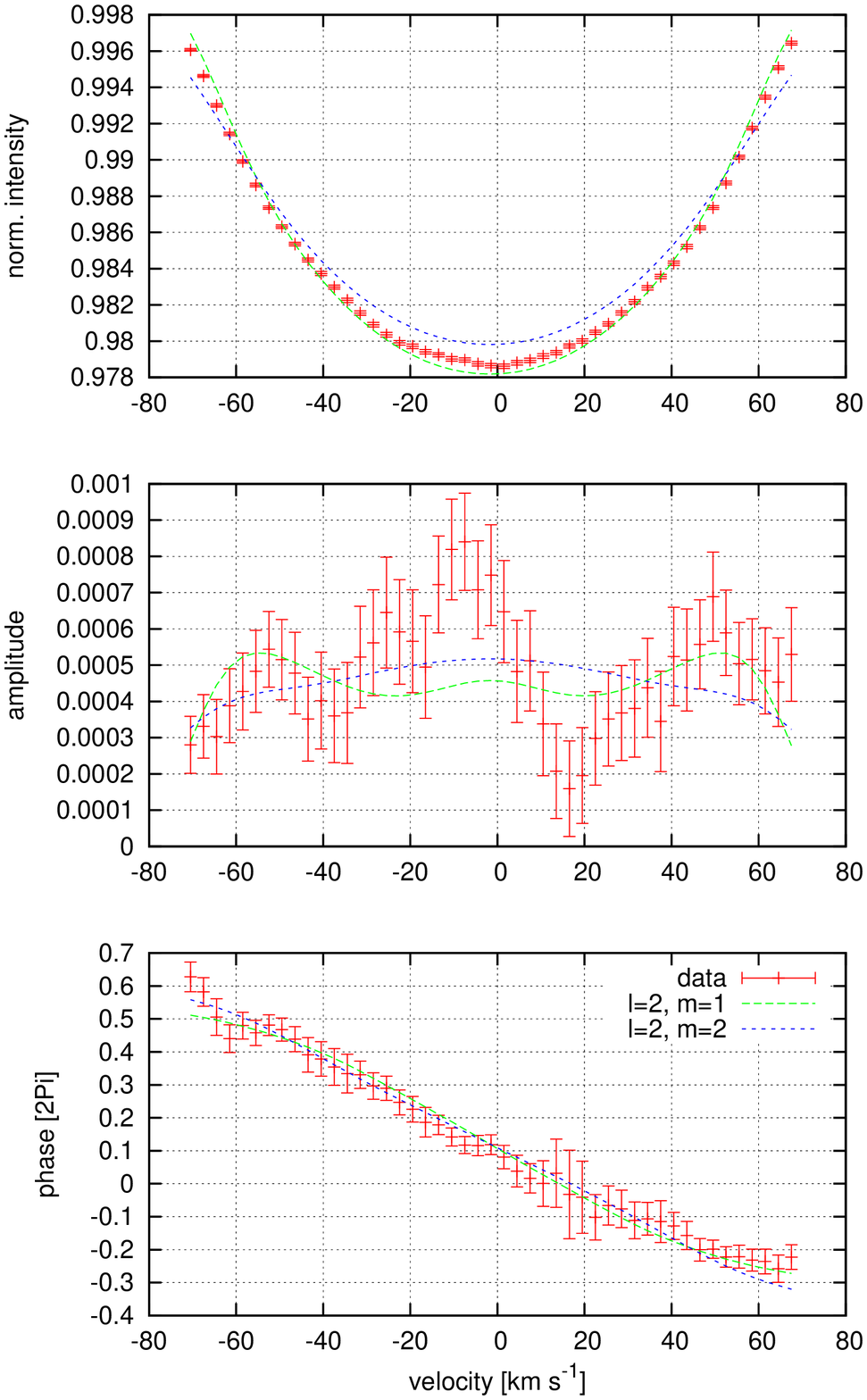}&
  \includegraphics*[width=55mm,bb=50 50 490 764,clip,angle=0]{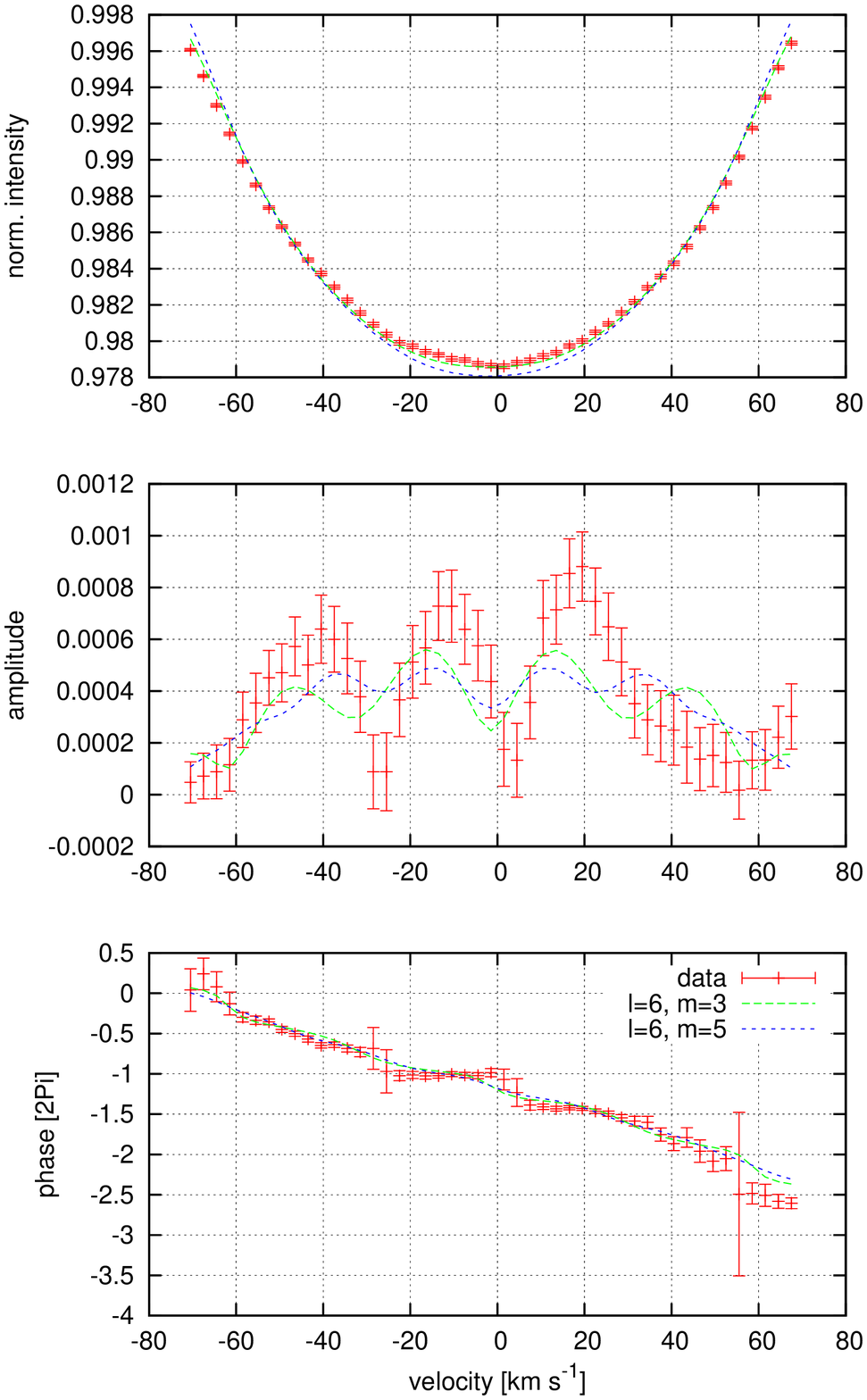}&
  \includegraphics*[width=55mm,bb=50 50 490 764,clip,angle=0]{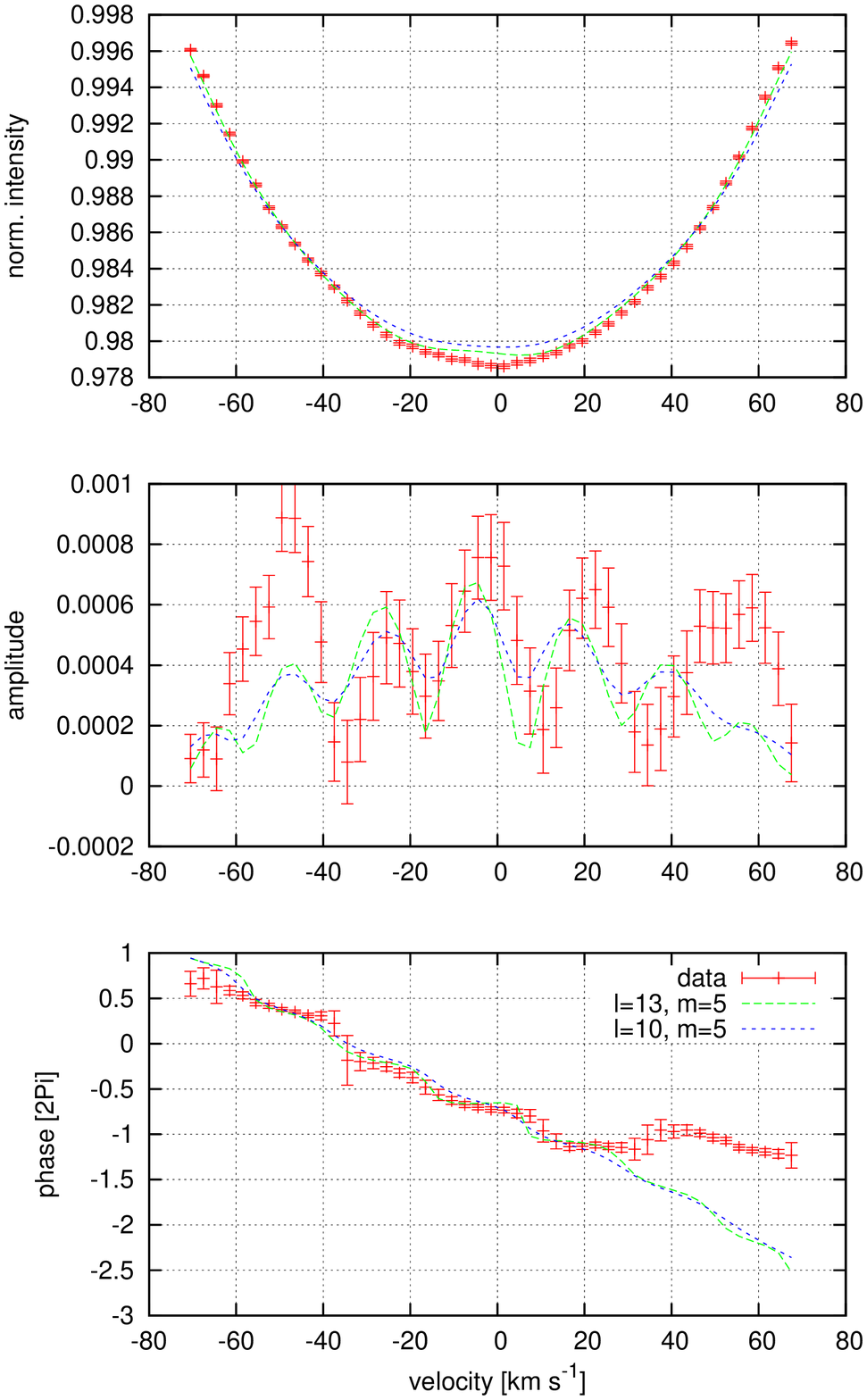}\\
\end{tabular}
\end{center}
\caption{Fourier parameters across the line profile of the detected pulsation frequencies. Each panel shows from top to bottom: zero-point, amplitude and phase. For both frequencies, the two best fits to the observed zero-point, amplitude in units of the continuum and the phase distribution in units of $2\pi$ are shown. The uncertainties are the formal errors computed from the least-squares fitting. Secondary component of RS Cha.}
\label{fig:fitzap2}
\end{figure*}

We carried out a mode identification with the FPF method for the frequencies $f_1$, $f_{\mathrm 2b}$, and $f_3$ and searched the same parameter space as for the primary component. The results are listed in Table~\ref{tab:modeid2}. According to our mode identification, $f_1$ is a low-degree pro-grade mode, whereas the two other higher frequencies are high-degree pro-grade oscillation modes. In comparison to the mode identification of the primary, the values of the goodness of fit ($\chi^2_\nu$) are higher for $f_{\mathrm 2b}$ and $f_3$, which implies a lower reliability of the derived identification.  For these two frequencies, the degree, $\ell$, is much better constrained than the value of $m$.
Figure~\ref{fig:fitzap2} shows the Fourier parameters of the observations and the two best fitting models for each mode.

\begin{table}[!ht]
\centering
\normalsize
\caption{Mode parameters derived from the application of the FPF method. For each pulsation mode, the four best solutions are shown. $\chi^2_\nu$ is the reduced chi-square value, $a$ is the velocity amplitude in~\kms;  \vsini\ is the projected rotational velocity; $\sigma$ is the width of the intrinsic Gaussian profile, both expressed in \kms. The inclination value was fixed at 83.4$^\circ$. Secondary component of RS Cha.}

\begin{tabular}{cccccc}
\hline
$\chi^2_\nu$ & $\ell$ & $m$ & $a$  &  \vsini & $\sigma$ \\  \hline
\multicolumn{6}{c}{$f_1=12.81$~\cd} \\ \hline
1.03 & 2 & 1  & 20.2 & 69 & 7.5\\
1.06 & 2 & 2  & 3.7  & 74 & 15.0\\
1.34 & 1 & 1  & 3.7  & 69 & 6.6\\
1.35 & 3 & 2  & 18.9 & 74 & 15.0\\
1.91 & 3 & 3  & 3.1  & 74 & 15.0\\
\hline\hline
\multicolumn{6}{c}{$f_{\mathrm 2b}=19.11$~\cd} \\ \hline
4.06 & 13 & 5  & 22.6  & 72 & 7.5\\
4.23 & 10 & 5  & 15.6  & 74 & 7.5\\
4.30 & 9  & 4  & 12.3  & 70 & 4.7\\
4.33 & 12 & 4  & 20.3  & 68 & 6.6\\
4.45 & 10 & 10 & 1.0   & 71 & 7.5\\
\hline\hline
\multicolumn{6}{c}{$f_3=24.56$~\cd}\\ \hline
2.09 & 6 & 3  & 20.9 & 70 & 7.5\\
2.15 & 6 & 5  & 7.7  & 69 & 5.6\\
2.21 & 7 & 4  & 28.6 & 71 & 11.2\\
2.34 & 7 & 6  & 13.7 & 74 & 11.2\\
2.34 & 6 & 6  & 1.6  & 69 & 11.2\\

\hline\hline

\hline
\end{tabular}
\label{tab:modeid2}
\vspace*{5mm}
\end{table}

\subsubsection{Results with the Fourier Fourier 2D method}

\begin{figure}[!ht]
  \includegraphics[width=70mm]{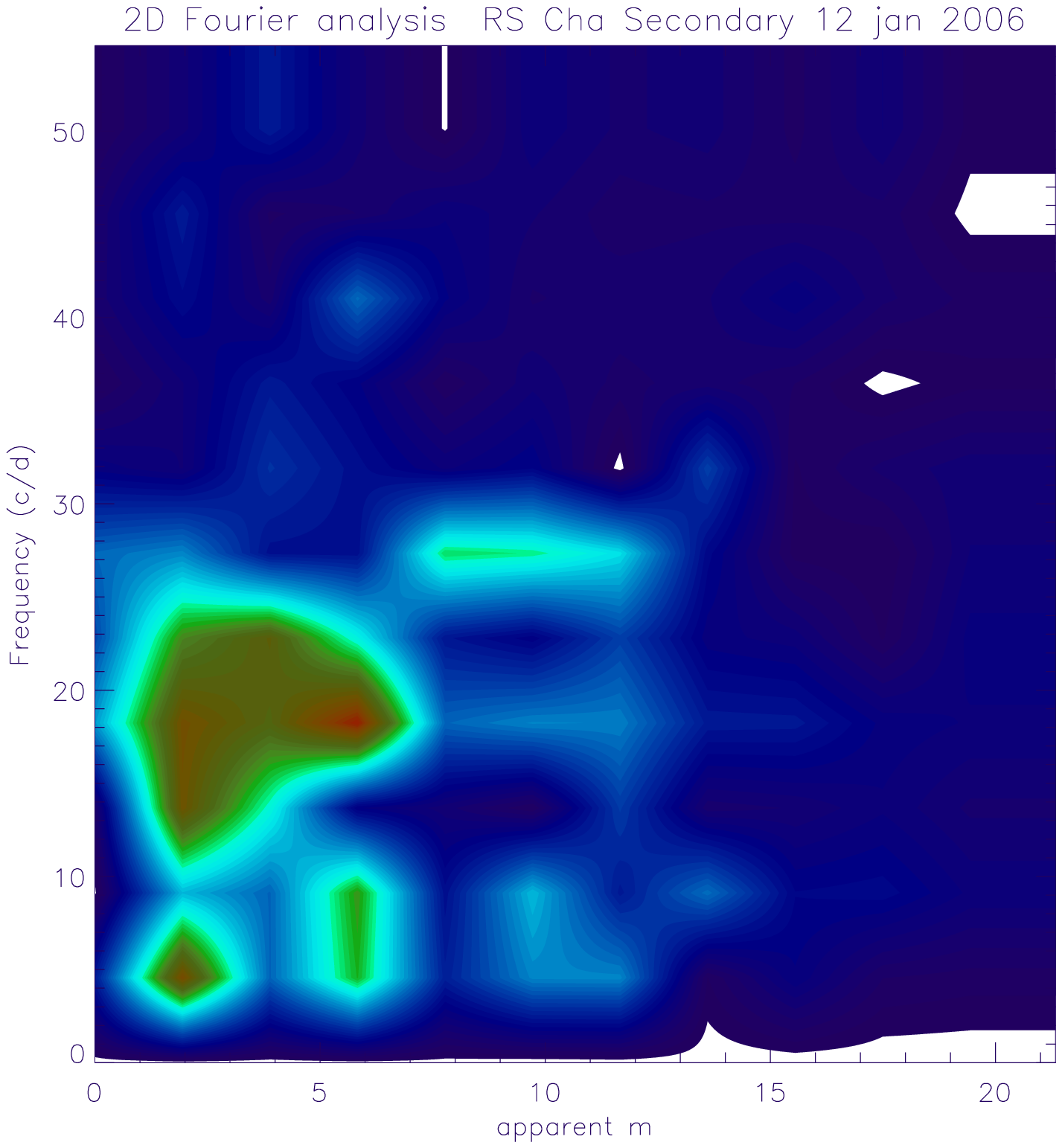}
  \includegraphics[width=70mm]{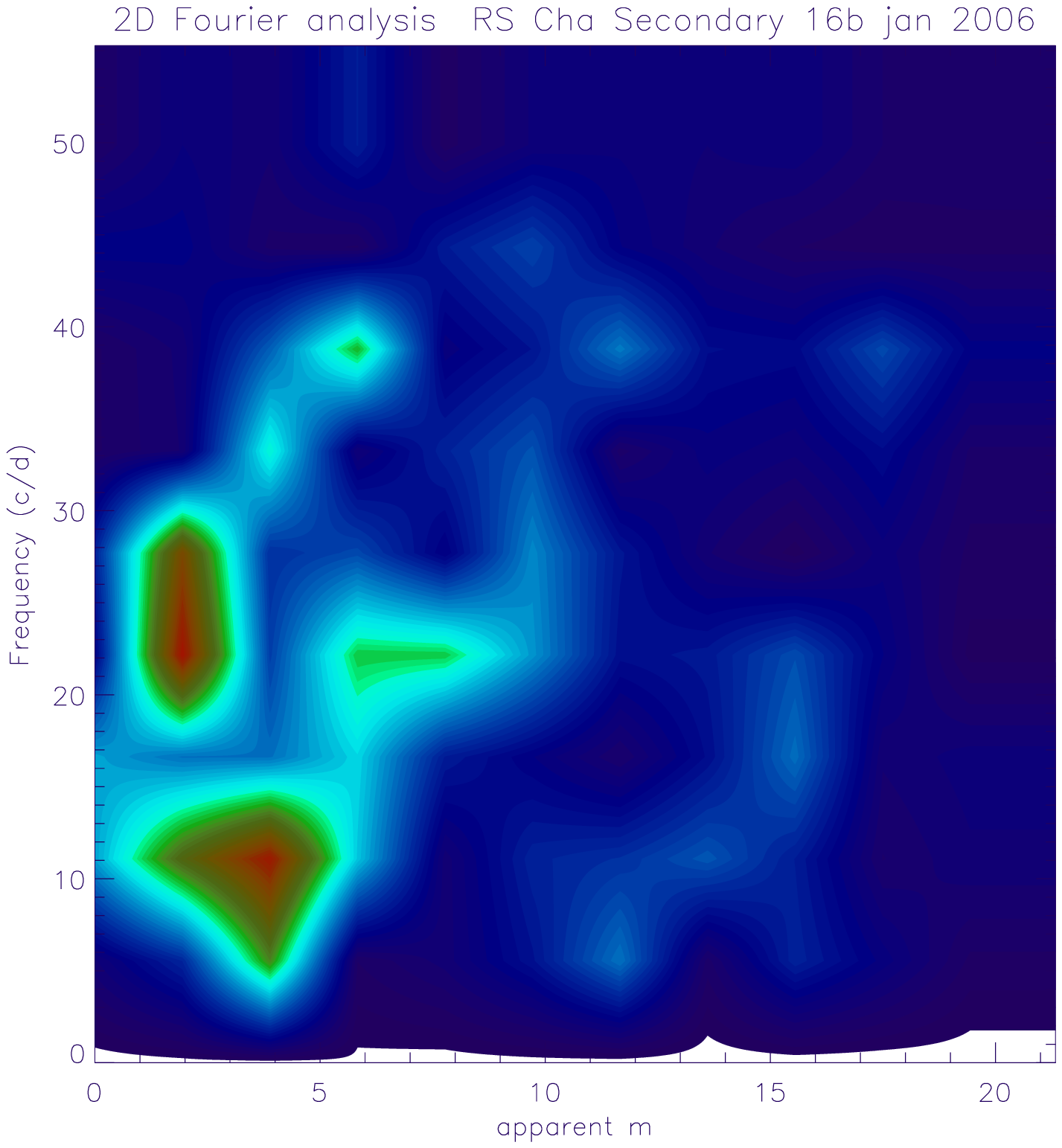}
  \includegraphics[width=70mm]{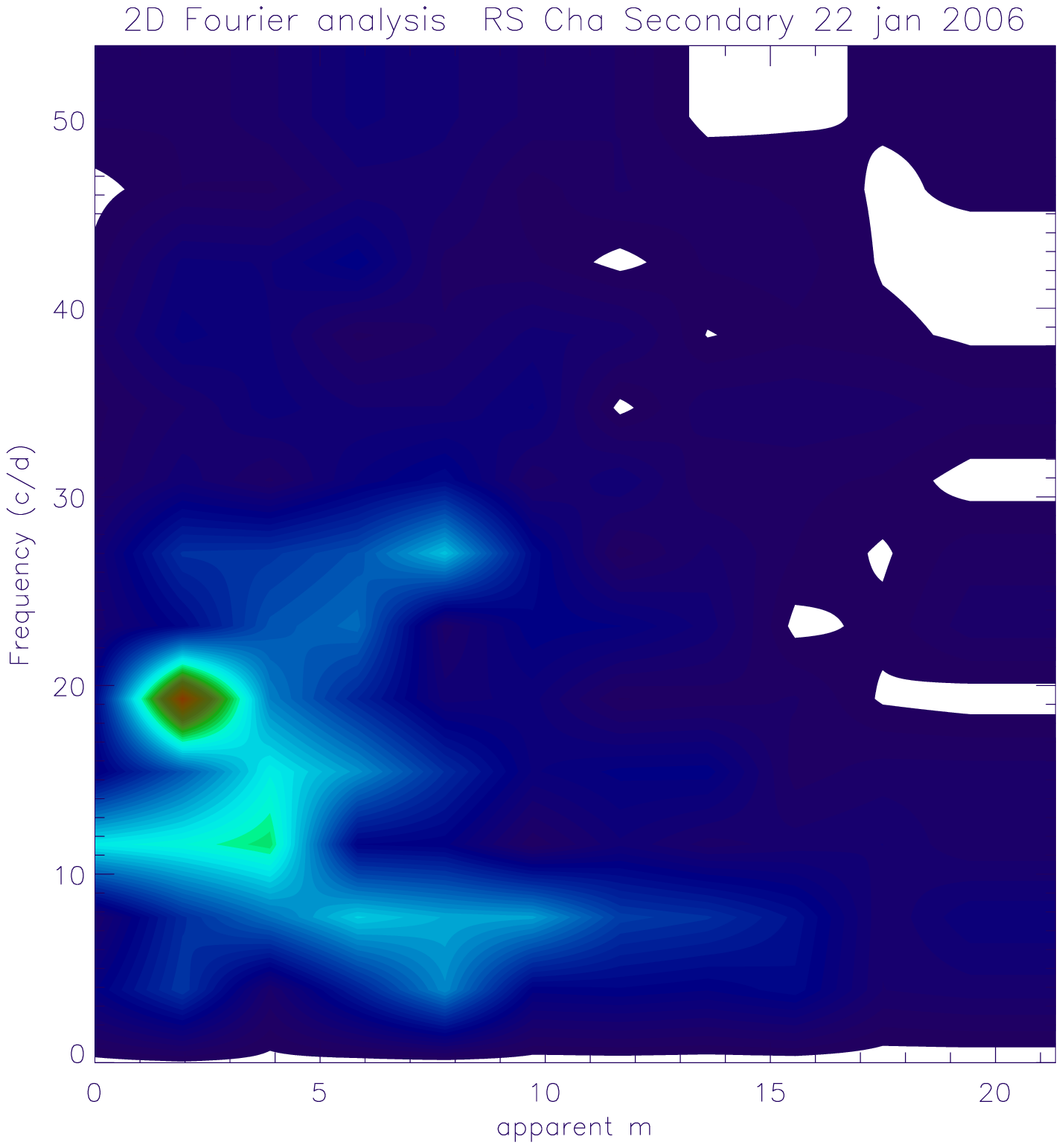}
\caption{F2D analysis on the Secondary component of RS Cha for the nights of Jan 12$^\mathrm{th}$, 16$^\mathrm{th}$ and 22$^\mathrm{nd}$ 2006.   }
\label{fig:f2D_Secondary}
\end{figure}

\begin{figure}[!ht]
  \includegraphics[width=70mm]{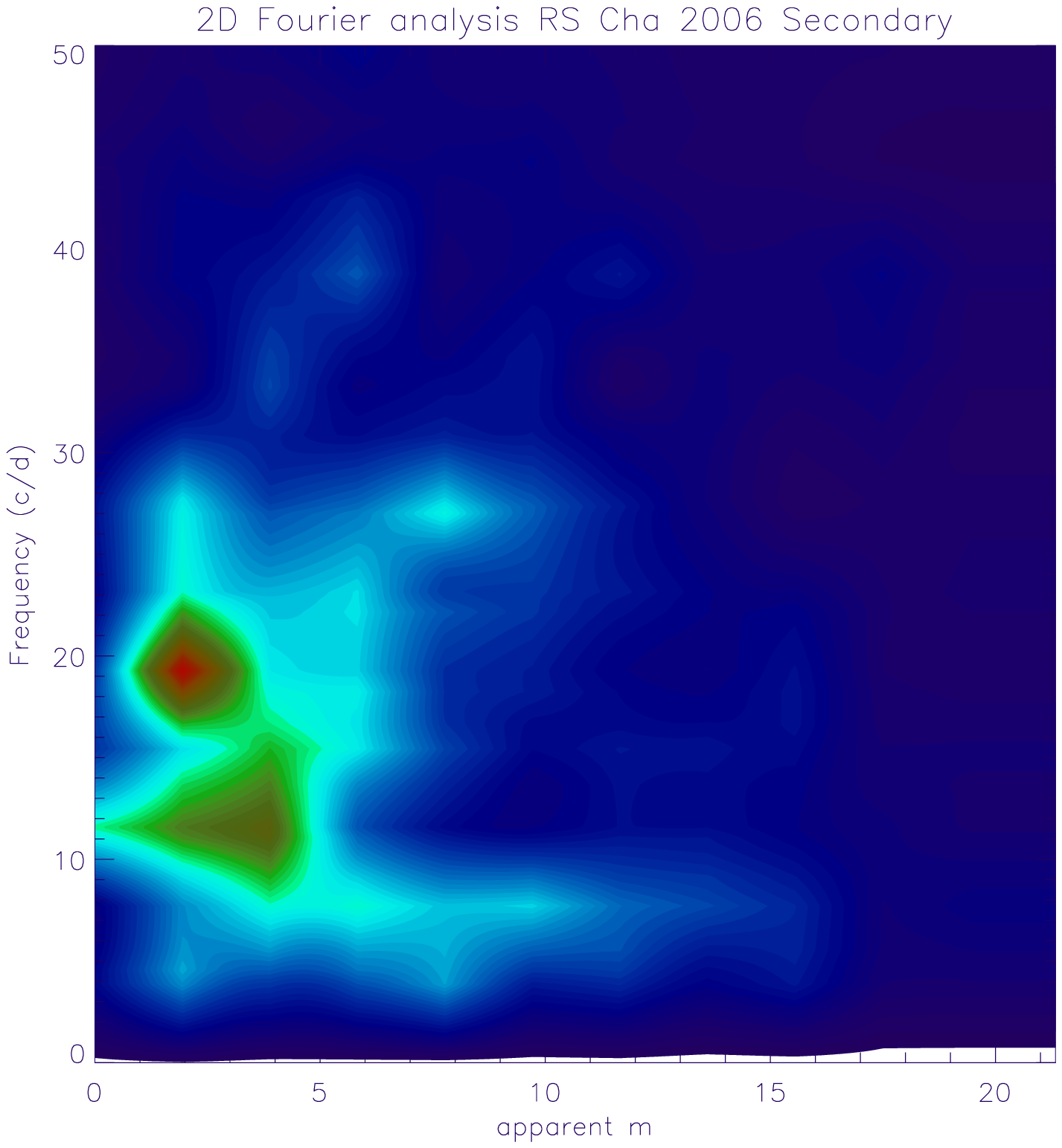}
\caption{Weighted F2D periodogram of the Secondary component of RS Cha for the nights of Jan 12$^\mathrm{th}$, 16$^\mathrm{th}$ and 22$^\mathrm{nd}$ 2006.}
\label{fig:f2D_Secondary_weighted}
\end{figure}

Figure~\ref{fig:f2D_Secondary} shows the nightly F2D analysis for the nights of 12$^\mathrm{th}$, 16$^\mathrm{th}$ and 22$^\mathrm{nd}$ of January 2006. The weighted F2D periodogram in Figure ~\ref{fig:f2D_Secondary_weighted} clearly exhibits
the presence of 2 main frequencies around 12\,d$^{-1}$ and 19\,d$^{-1}$, which confirms the frequency determination by the FPF method for $f_1$ and $f_{\mathrm 2b}$. The F2D method indicates an $\ell$ value of 2 or 3 for $f_1$ and confirms the FPF determination (within the error bars). However, the F2D method attributes a low $\ell$ value of 0 to $f_{\mathrm 2b}$ and identifies therefore this mode as radial, quite in contradiction with the FPF result for the same frequency, indicating a high degree $\ell$ =  9-13. 

Table \ref{tab:fou2dsecondary} summarizes the results for the primary component based on the F2D analysis.

\begin{table}
\caption[]{Frequency search and mode identification. Summary of the weighted F2D analysis on the primary component. 
Columns are as following: (1) frequency, (2) possible identification with FPF result, (3) apparent $\left | m \right |$, (4) expected $\ell$ value.\\}
\begin{tabular}{cccc}
\hline\hline

       d$^{-1}$        &  FPF frequency & app.  $\left | m \right |$ & $\ell$\\\hline
         19.                 &   $f_{\mathrm 2b}$		& 2	& 0 \\
         12. 		  &   $f_1$			& 4    & 2 or 3 \\  \hline         
\end{tabular}
\label{tab:fou2dsecondary}
\end{table}

\section{Discussion and conclusions}
\label{dissconc}

For the first time we detected by direct spectroscopic means non-radial pulsations in both components of a binary Herbig Ae star. 
We identified  with a high level of confidence in the profile variations of the primary  and secondary component two, respectively three pulsation frequencies.
A first identification by two complementary methods revealed very different main pulsation modes for the two components: while the dominant mode of the primary component seems to be a high degree pro-grade mode with $\ell$ = 10 or 11, the dominant mode of the secondary component is identified as a low degree mode with $\ell$ = 0,1 or 2. Two different methods have been applied to perform a mode identification of both components. 
The Fourier 2D method as presented by \cite{kennelly} provides direct insights on the spatial structure of the oscillation. However, the derived apparent $\left | m \right |$ value seems to be linked to $\ell$ in the way described here-above, but more detailed simulations for different configurations have to be carried out in order to understand potential biases in the degree $\ell$ determination. We therefore privilege the $\ell$ determination as provided by the FPF method.
Table \ref{mainresults} summarizes the main results for the dominant pulsation modes of each component.
As can be seen, both methods yield very different results on the determination of the degree $\ell$ of $f_{\mathrm 2b}$ of the secondary component, perhaps due to an insensibility of the F2D method to tesseral mode signatures.

\begin{table}
\caption[]{Frequency search and mode identification. Summary of the main results for the dominant frequencies of each component. The ($\ell$,m) couples indicated for the FPF method correspond to the best identification, in decreasing order. For each of the frequencies detected by the FPF method, we identified an $\ell$ value as detected by the F2D method. \\}
\begin{tabular}{cccc}
\hline\hline
 \multicolumn{4}{c}{\bf Primary component} \\ 
 &&&\\
               & d$^{-1}$ & Identification FPF  & Identification F2D \\\hline
               &                 &     best ($\ell$,m)    &   \\
  ~~~~~ $f_1$ ~~~~~~ &   21.11 & (11,11), (11,9), (10,6) &  $\ell$ = 8-10\\ 
   $f_2$  & 30.38  &   (10,6), (9,5), (11,7) & \\\hline 
   &&&\\\hline\hline
 \multicolumn{4}{c}{\bf Secondary component} \\ 
 &&&\\
               & d$^{-1}$ & Identification FPF  & Identification F2D \\\hline
                              &                 &     best ($\ell$,m)    &   \\
   $f_1$  &  12.81  &  (2,1), (2,2) &  $\ell$ =  2 or 3\\ 
   $f_{\mathrm 2b}$  &  19.11  & (13,5), (10,5) &  $\ell$ = 0\\
   $f_3$  &  24.56  & (6,3), (6,5) & \\   \hline \hline
\end{tabular}
\label{mainresults}
\end{table}

The precise redetermination of the binary orbit reveals a slight gradient change of the observed - calculated phase shift curve, which might be the 
signature of changes in the orbital period of the order of  $\Delta P$/P = (7.5 $\pm$ 0.7) 10$^{-6}$. This tendency, if confirmed, might be due to the presence of a third body in the RS\,Cha system or due to intrinsic orbital period changes, which then need to be explained.

Next steps require a precise redetermination of the fundamental parameters and a search for the intrinsic rotation periods of both components. The first frequencies are now known and their associated modes are constrained by the present work. In subsequent work, a numerical model will be developed in order to constrain the internal structure of both components.

\begin{acknowledgements}
We want to thank the University of Canterbury, New Zealand, for providing us access to their Mt. John 1m McLellan telescope, but also to the staff of the Mt. John observatory for supporting us during the observing run. WZ is supported by the FP6 European Coordination Action HELAS and by the
Research Council of the University of Leuven under grant GOA/2003/04.
\end{acknowledgements}

\end{document}